\documentclass[aps,groupedaddress,preprint,superscriptaddress,showpacs]{revtex4-1}

\usepackage[T1]{fontenc} 
\usepackage[portuguese,english]{babel}
\usepackage{hyperref}
\usepackage{amsmath,bm}
\usepackage{xfrac} 
\usepackage{mathtools}
\usepackage{booktabs}
\usepackage{xcolor}
\usepackage[normalem]{ulem}
\usepackage{xr}
\usepackage{floatrow}
\usepackage[position=top]{subfig}
\usepackage[version=4]{mhchem}
\usepackage{booktabs}
\usepackage{etex}
\usepackage{morefloats}
\usepackage{siunitx}

\vfuzz \hfuzz
\hypersetup{colorlinks=true,citecolor=blue,urlcolor=blue}

\begin{document}

\title{Electronic and optical properties of two-dimensional flat band triphosphides}

\author{Gabriel Elyas Gama Araújo}
\affiliation{Institute of Physics, Federal University Goi\'as, Campus Samambaia, 74690-900, Goi\^ania, Goiás, Brazil.}

\author{Lucca Moraes Gomes}
\affiliation{Institute of Physics, Federal University of Goi\'as, Campus Samambaia, 74690-900, Goi\^ania, Goiás, Brazil.}

\author{Dominike Pacine de Andrade Deus}
\affiliation{Instituto Federal de Educação, Ciência e Tecnologia do Triângulo Mineiro,
Uberlândia, $38400-970$, Minas Gerais, Brazil}
\email{dominike@iftm.edu.br}

\author{Alexandre Cavalheiro Dias}
\affiliation{Institute of Physics and International Center of Physics, University of Bras{\'{i}}lia, Bras{\'{i}}lia $70919$-$970$, Distrito Federal, Brazil}
\email{alexandre.dias@unb.br}

\author{Andr\'eia Luisa da Rosa}
\email{andreialuisa@ufg.br}
\affiliation{Institute of Physics, Federal University Goi\'as, Campus Samambaia, 74690-900, Goi\^ania, Goiás, Brazil.}

\begin{abstract}
 In this work we use first-principles density-functional theory (DFT)
 calculations combined with the maximally localized Wannier function
 tight binding Hamiltonian (MLWF-TB) and Bethe-Salpeter equation (BSE)
 formalism to investigate quasi-particle effects in 2D electronic and
 optical properties of triphosphide based two-dimensional materials
 XP$_3$ (X =  Ga, Ge, As; In, Sn, Sb; Tl, Pb  and Bi). We find that with
 exception of InP$_3$, all structures have indirect band gap.  A
 noticeable feature is the appearance of flat valence bands associated
 to phosphorous atoms, mainly in InP$_3$ and GaP$_3$ structures. Furthermore,
 AIMD calculations show that 2D-XP$_3$ is stable at room temperature,
 with exception of TlP$_3$ monolayer, which shows a strong distortion
 yielding to a phase separation of the P and Tl layers. Finally, we show that
 monolayered XP$_3$ exhibits optical absorption with strong excitonic
 effects, thus revealing exciting features of these monolayered
 materials.
 
\end{abstract}
\maketitle

\section{Introduction}

Two-dimensional black phosphorus (phosphorene) has recently attracted
attention due to its tuneable band gap and promising applications in
electronic and optoelectronic
devices\,\cite{25,26,27,47,x6,x8,li2014black,Chaudhary_2022}. Phosphorene
has been of particular interest since it has a direct electronic band
gap, which varies from 0.3 (bulk) to 2.0 eV
(monolayer)\,\cite{Yang2015}. Black phosphorene shows a quasi-particle
optical gap of 2.2\,eV, with a near-band-edge recombinations are
observed at 2\,K with excitonic transitions at 0.276 eV and 0.278
eV\,\cite{Carre_2021,NNano15}.

Beyond monoelemental low dimensional phosphorous, other combinations
of phosphorene with group-III, GaP$_3$
\cite{sbp3_and_gap3_ph1,In-Ga-Sb-Sn-triphos_ph}, InP$_3$
\cite{In-Ga-Sb-Sn-triphos_ph}, group-IV,
GeP$_3$\,\cite{gep3_cr_pacine,gep3_ph,Pb-Ge-Sn_ph},
SnP3\cite{Pb-Ge-Sn_ph,In-Ga-Sb-Sn-triphos_ph}, PbP3\cite{Pb-Ge-Sn_ph}
and group V AsP$_3$ \cite{NJP2024},
BiP$_3$\cite{Langmuir2023,PCCP2021,ASS2020,NL2017,ASS2024,MSSP2024,bip3_ph1}
and SbP$_3$\cite{sbp3_and_gap3_ph1,In-Ga-Sb-Sn-triphos_ph}.

Theoretical calculations predict SnP$_3$ monolayer to have an indirect
electronic bandgap of 0.43\,eV, which can be tuned by external
strain\,\cite{pssb2023}. InP$_3$, GaP$_3$, SbP$_3$ and SnP$_3$ have
been theoretically investigated and found promising for low thermal
conductivity\,\cite{Nanoscale2020}. Finally, the photocatalytic
properties of AlP$_3$ and GaP$_3$ for water splitting and hydrogen
production have been addressed\,\cite{ACS2020}.

In systems with finite band width, electrons can be confined in real
space in crystals which possess the so-called flat bands in momentum
space. Examples include f-electron systems with Kondo effect and heavy
fermions\,\cite{Science_Si_2010}, fractional quantum Hall
effect\,\cite{Tsui_1982}, and twisted bilayer graphene superlattices
which show unconventional superconductivity\,\cite{Cao_Nature_2018}

Until now, there has been scarse experimental data on XP$_3$
structures\,\cite{InP3_exp}. Therefore, in this work we gather
electronic, and optical properties to reveal the main feature of this
important class of large gap materials under the same level of
approximation.  We find that with exception of InP$_3$, all structures
have indirect band gap.  We show that monolayered XP$_3$ exhibits
optical absorption with strong excitonic effects. The exciton binding
energy is significantly large for X = Ga, Tl, Ge, Sn and Pb. In particular,  \ce{InP3}, \ce{GaP3} and \ce{BiP3} shows a good solar harvesting efficiency around \SI{20}{\percent} - \SI{30}{\percent}, being attractive for solar cell applications.

\section{Computational details} 

We have performed first-principles calculations within the GGA (generalized-gradient approximation according to the  parameterization of PBE (Perdew-Burke-Ernzerhof) \,\cite{pbe} and HSE (Heyd-Scuseria-Ernzenhof)\,\cite{hse} to describe the exchange an correlation potential as implemented in the Vienna \textit{ab initio} Simulation Package (VASP).\cite{Kresse_13115_1993, Kresse_11169_1996}. The electronic wave-functions were built using the projected augmented wave (PAW) method \cite{paw}.  The single-electron Kohn-Sham wave-functions were expanded in plane-waves up to the energy cutoff of 400\,eV. 

Structure optimizations ensured that the  forces on the atoms  were below 0.01 eV/{\AA}. In order to eliminate any spurious interactions between the monolayer and its periodic images in the $z$-direction, we incorporated a vacuum layer with a thickness of 16\,{\AA} in each monolayer unit cell. For all calculations, \textbf{k}-meshes were automatically generated utilizing the Monkhorst-Pack scheme \cite{monkhorst1976special}. Electronic structure calculations using (5x5x1) {\bf k}-points show negligible difference to results using a (9x9x1) {\bf k}-point sampling.


Ab-initio molecular dynamics (AIMD) calculations were performed using
a ($3\times3\times1$) supercell with a $(2\times2\times1)$ k-points
sampling at T = 300K and a NVT ensemble. The Andersen thermostat was
coupled to the XP$_3$ monolayers.  Simulation times between of 10\,ps
with time steps of 5\,fs have been performed. In order to calculate
the optical properties, we solve the Bethe-Salpeter equation (BSE) by
employing the the Coulomb truncated 2D potential
(V2DT),\cite{Rozzi_205119_2006}. Based on that, the linear optical
response of the materials was computed to include excitonic effects by
solving the Bethe–Salpeter equation (BSE)\cite{Salpeter_1232_1951} by
employing the WanTiBEXOS package\cite{Dias_108636_2022}. We first
created a maximally localized Wannier function tight-binding (MLWF-TB)
Hamiltonian to conduct these evaluations derived from DFT-HSE06
calculations including SOC through the Wannier90
package,\cite{wannier90}Our calculations were performed with a
120{\AA} density of \textbf{k}-points to determine the real and
imaginary parts of the dielectric function. The optical properties
were calculated at the Independent Particle Approximation (IPA) and
BSE levels, considering the necessary number of conduction and valence
bands to describe the optical properties in the solar emission region
(i.e \SIrange{0}{4}{\electronvolt}).\cite{Astm_1_2012} Additional
information concerning the BSE parameters is available in the SI
section S3.

The solar harvesting efficiency of these monolayers was estimated
through the power conversion efficiency (PCE), considering the AM1.5G
model for the solar emission spectrum,\cite{Astm_1_2012} employing the
Shockley--Queisser limit (SQ-limit)\cite{Shockley_510_1961} and the
spectroscopy limited maximum efficiency (SLME)
method.\cite{Yu_068701_2012} These simulations considers the solar
cell operating at \SI{300}{\kelvin}. The absorbance, used for the PCE
estimative at SLME, was evaluated using the total absorption
coefficient, obtained at BSE or IPA levels, from the summation of the
dielectric function diagonal components. In our simulations, we also
assumed the \ce{XP_3} monolayer thickness equals to the material
thickness plus de van der Waals (vdW) length (i.e
\SI{3.21}{\angstrom}),\cite{Bernardi_3664_2013} detailed in SI section
S5. The addition of vdW length to \ce{XP_3} monolayer thickness is
justified in the work of Bernardi et. al.,\cite{Bernardi_3664_2013}
where this procedure was shown necessary to estimate graphene
absorbance due its atomic layer thickness, in order to reach results
closer to experimental measures. Production of images have been provided by the VESTA\,\cite{VESTA} and grace packages\,\cite{xmgrace}.

\subsection{Structural properties}

\begin{figure}[H]
\includegraphics[width=7cm]{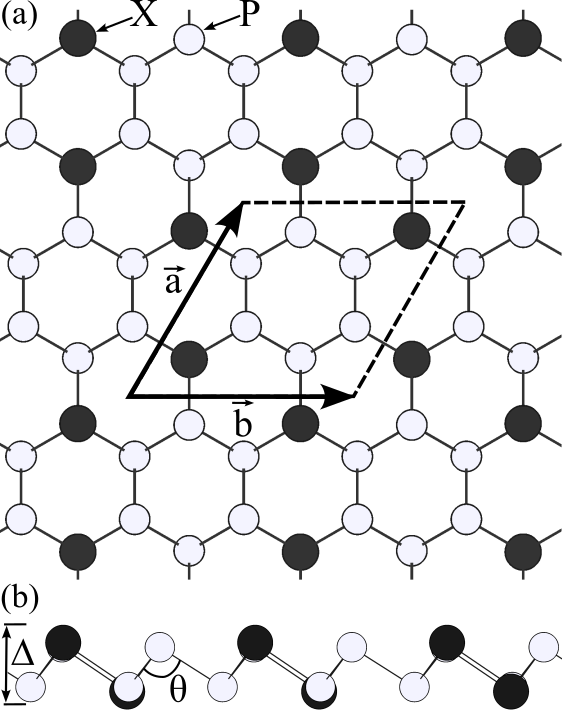}
\caption{\label{fig:geometries-monolayers}  a) Top view and b) side view of hexagonal monolayered XP$_3$ (X =  Ga, Ge, As; In, Sn, Sb; Tl, Pb  and Bi). The lattice parameter $a$, $\theta$ and the buckling $\Delta$.}
\end{figure}

Monolayered XP$_3$ has a honeycomb structure in which each X atom
forms three X-P bonds with three adjacent P atoms. Each P atom forms
two P-P bonds and one X-P bond, respectively, as shown in
Fig.\,\ref{fig:geometries-monolayers}.  Table
\ref{tab:geometry_formation} shows the optimized lattice parameters
for groups III-, IV- and V-P$_3$. It is possible to see that group III
and group V, have similar buckling $\Delta$, with exception of TlP$_3$. The enthalpy of formation at T = 0\,K of the investigated XP$_3$ compounds was calculated as:

\begin{equation}
\Delta H_f^{\rm T=0K} {(\rm XP_3)} = E_{\rm XP_3} - E_{\rm X} - E_{\rm BP-bulk},
\end{equation}

\noindent where E$_{\rm XP_3}$, E$_{\rm X}$ and E$_{\rm P}$ are the
total energies of monolayered XP$_3$, monoelemental bulk metal X and
bulk black phosphorous at both GGA-PBE and HSE levels,
respectively. The results are shown in Table
\ref{tab:geometry_formation}. As a general behavior, most of the compounds have formation enthalpies
between 1.0 and 2.0 eV. In particular for AsP$_3$ a very small value is found.

\begin{table}[H]
\centering
\begin{tabular*}{12cm}{@{\extracolsep{\fill}}l ccccccc}
\hline
& & $a$(\AA) & $\Delta$ (\AA)  &  P-P (\AA)  & X-P(\AA)   & X-X (\AA) & $\Delta H^{T=0K}_f$ (eV)\\  
\hline            
& GaP$_3$& 7.19 & 1.21  & 2.23 & 2.23 & 4.15 & 0.71 \\ 
Group III & InP$_3$& 7.53 & 1.22  & 2.23 & 2.56 & 4.35 & 1.34\\ 
&TlP$_3$& 7.18 & 2.34 &  2.19 & 3.11 & 4.76 & 2.32 \\ \hline 
&GeP$_3$& 6.95 & 2.38 & 2.17 & 2.50 & 4.67 & 1.62  \\ 
Group IV &SnP$_3$& 7.15 & 2.84 &  2.17 & 2.70 & 5.01 & 1.14 \\ 
&PbP$_3$& 7.28 & 2.92 & 2.17 & 2.79 & 5.12 &   1.13   \\ \hline
&AsP$_3$& 6.72 & 1.48 &  2.25 & 3.39 & 4.16 & 0.16\\ 
Group V &SbP$_3$& 7.00 & 1.79 &  2.25 & 2.60 & 4.42 & 0.53 \\ 
&BiP$_3$& 7.14 & 1.91 &  2.24 & 2.70 & 4.55 & 0.63  \\ \hline
\end{tabular*}
\caption{Lattice parameter $a$, buckling $\Delta$, bond lengths P-P, X-P and X-X and formation enthalpy $\Delta H^{\rm T=0K}_f$ for XP$_3$ (X =  Ga, Ge, As; In, Sn, Sb; Tl, Pb  and Bi) within GGA-PBE. }
\label{tab:geometry_formation}
\end{table}

\floatsetup[figure]{style=plain,subcapbesideposition=top}

\subsection{Electronic properties}

According to the periodic table of elements, the electronegativity of should follow the trend from left to right: Ga $<$ Ge $<$ As; In $<$ Sn $<$ Sb; Tl $<$ Pb $<$ Bi and from top to bottom: Ga $>$ In $>$ Tl; Ge $>$ Sn $>$ Pb; As $>$ Sb $>$ Bi. The larger the electronegativity, the smaller the band gap. For group III and V this is very clear according to Table \ref{tab:gap}. On the other hand, group IV -P$_3$ does not seem to follow this trend. HSE improves the gap by around 50\,\% in most cases. In particular, for GeP$_3$ this value almost doubles at HSE compared to calculations at GGA-PBE level. For group IV, as the buckling $\Delta$  increases, the electronic band gap also increases. Finally for group V, as the buckling increases, the  band gap decreases for both GGA-PBE and HSE functionals. Therefore, there is an interplay between ionicity, band gap and buckling.

For group IV, as the buckling $\Delta$  increases, the electronic band gap follows this behavior and also increases. Finally for group V, as the buckling increases, the  band gap decreases for both GGA-PBE and HSE functionals, suggesting  an interplay between ionicity, band gap and buckling.
 
\begin{figure}[H]
\includegraphics[width=15.2cm]{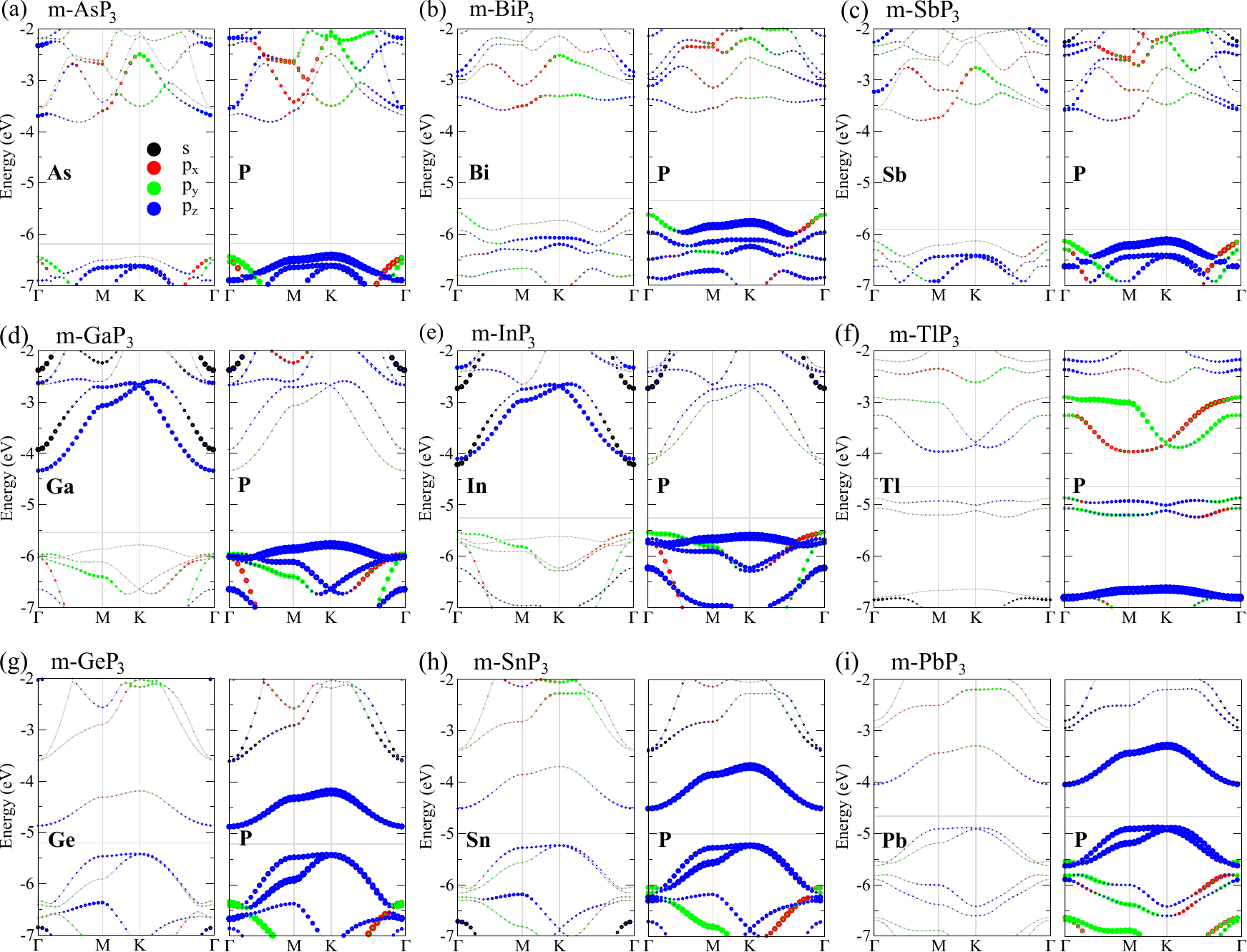}
\caption{\label{fig:bandas-monolayers} Orbital projected band structure of XP$_3$ compounds calculated within DFT-HSE06. Red: $p_x$, green: $p_y$ and blue: $p_z$. The size of the dots is proportional to the orbital contribution to the state. The horizontal dotted line is the work funcion value. All calculations include SOC.}
\end{figure}

\begin{figure}[H]
\includegraphics[width=15.2cm]{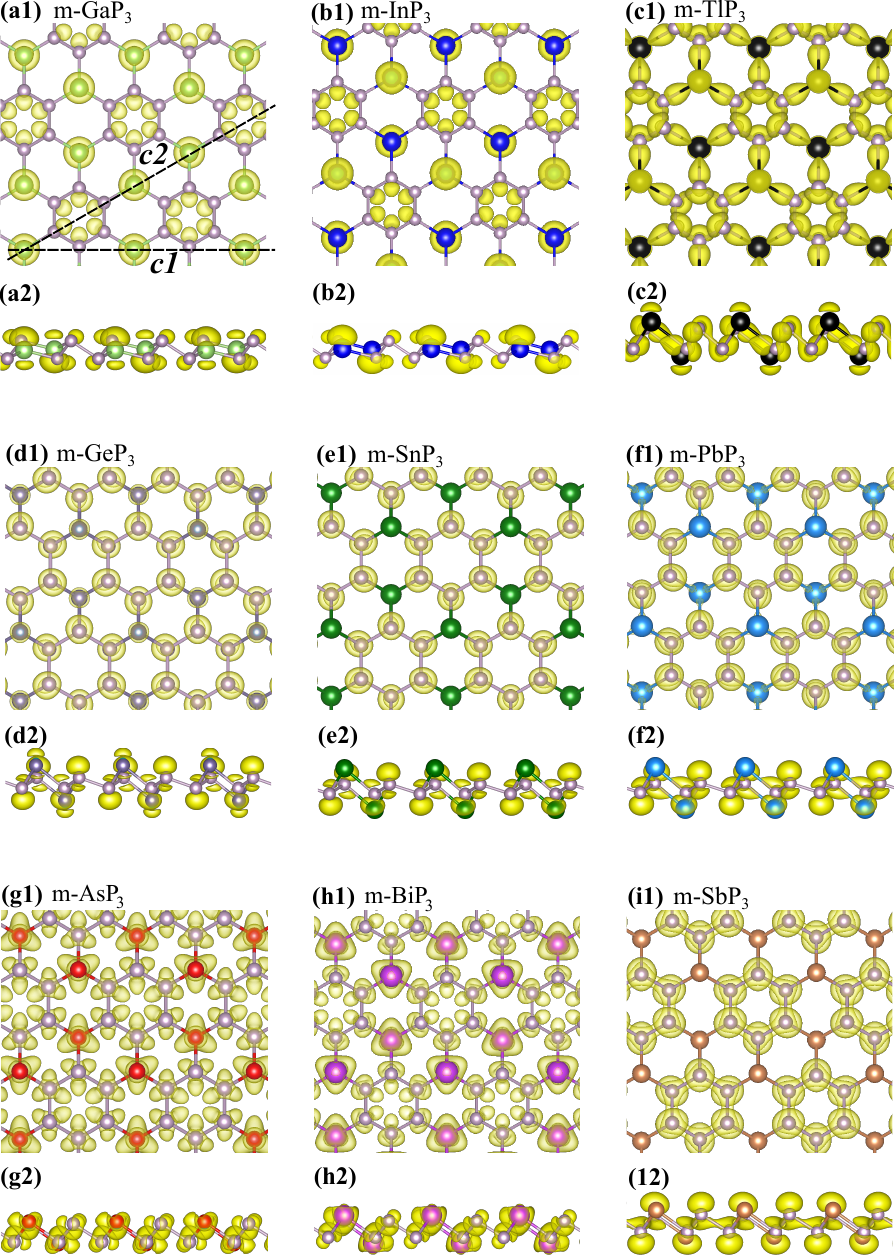}
\caption{\label{fig:lumo-monolayers} Projected charge at LUMO of XP$_3$ structures. The isosurface is equal to 0.002 e/\AA$^{3}$.}
\end{figure}

From Fig.\,\ref{fig:bandas-monolayers} with exception of InP$_3$, all
triphosphides have indirect gap. The top of the valence band has
majoritarily X-$p$ orbitals. At the K-point the major contribution
comes from the p$_z$ orbitals. As a general features, all layers have
flat band character, specially along the M-K direction. In flat band
materials the energy does not depend on crystal momentum and the
charge carriers have a zero group velocity and an infinite effective
mass. This feature could lead to several interesting properties, such
as ferromagnetism, superconductivity and topological
states\,\cite{hase_2018,kobayashi_2016}.  From
Fig.\,\ref{fig:lumo-monolayers} we see that the lowest unocuppied
molecular orbital (LUMO) is localized at the cation X-atom.

\begin{table}[H]
\centering
\begin{tabular*}{12cm}{@{\extracolsep{\fill}}l lccc}
\toprule
 & & \multicolumn{2}{c}{E$_{\rm gap}$ (eV)}  &  Work function (eV) \\ \hline
& & GGA-PBE & HSE & HSE\\  
\hline           
& GaP$_3$ & 0.78  & 1.45 (ind.) & 5.54 \\ 
Group 3A & InP$_3$ & 0.70   & 1.32 (dir.) & 5.25 \\ 
& TlP$_3$ &  0.57  & 0.90 (ind.) & 4.65\\ \hline
& GeP$_3$&  0.28  & 0.54 (ind.)  & 5.21       \\ 
Group 4A & SnP$_3$& 0.45  & 0.71 (ind.)  & 5.00 \\  
& PbP$_3$& 0.55  & 0.84 (ind.)   & 4.66         \\ \hline
& AsP$_3$& 1.89   & 2.61 (ind.)    & 6.19   \\ 
Group 5A & SbP$_3$& 1.64  & 2.34 (ind.) & 5.90\\ 
& BiP$_3$& 1.42 & 2.00  (ind.)   & 5.31     \\ \bottomrule
\end{tabular*}
\caption{Electronic band gap E$_{\rm gap}$  calculated within the GGA-PBE and HSE06 and work function  within HSE. All calculations include SOC.}
\label{tab:gap}
\end{table}

The work function is one critical parameter in electronic and optoelectronic devices. 
Work function and electron affinity are among the most important properties of semiconductors, which play essential roles in functional properties and device performance once interfaces or junctions are involved, such as metal-semiconductor junctions in devices or hetero/homojunctions for photovoltaic cells. 

Moreover, the role of the exchange-correlation functional was also addressed. Because PBE tends to underestimate the band gap, we have used the HSE functional with 25\% Hartree-Fock exchange to perform band structure calculations. The electronic band gap varies drastically from  from GGA-PBE to HSE. For example, the difference increases 27\% for AsP$_3$ and 48,\% for GeP$_3$, highlighting the importance of correct exchange-correlation treatment.

The work function is the energy needed to remove an electron from the surface of a solid to the vacuum  level. Here, we report a large modulation of the work function in XP$_3$ by changing the X atom.  In Table\,\ref{tab:gap} we can see that the work function varies from 4.65-6.19 eV. As matter of comparison, in XP$_3$, the small value we find is comparable to the lower limit of graphene work function, reported to be 4.30 eV\,\cite{bip3_pacine,wf_graphene}.  In other semiconductors, the work function for p-type (n-type) Si is 4.55-4.74 eV (4.63–4.66)\,eV. We therefore expect that even large work function modifications can be achieved in XP$_3$ materials by surface adsorption. 

In Fig.\,\ref{fig:lumo-monolayers} we can see that the charge is more  localized on the phosphor atom. 

\section{Molecular Dynamics simulations}

\begin{figure}[H]
\sidesubfloat[a]{\includegraphics[width=4cm]{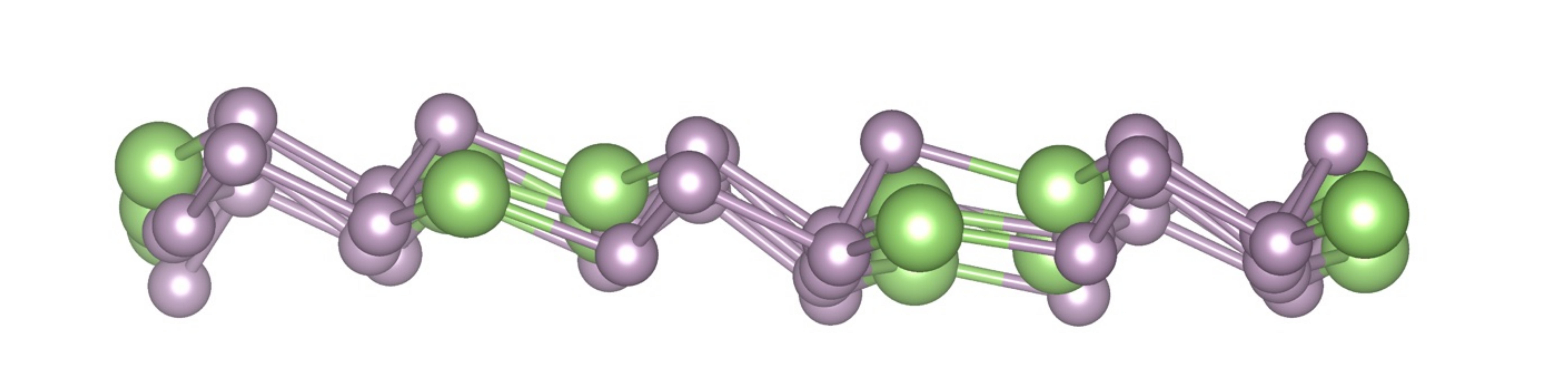}}
\sidesubfloat[b]{\includegraphics[width=4cm]{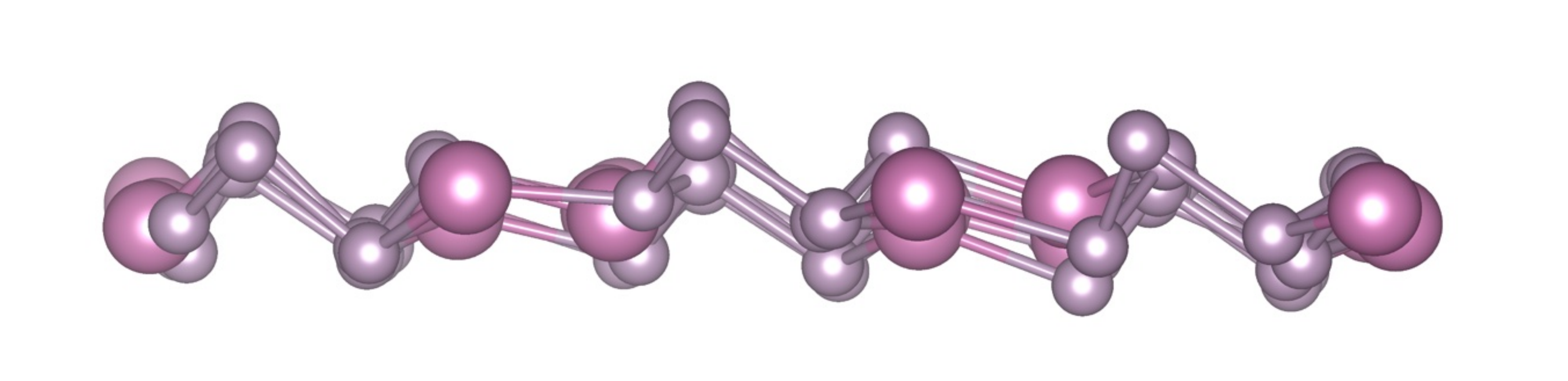}}
\sidesubfloat[c]{\includegraphics[width=4cm]{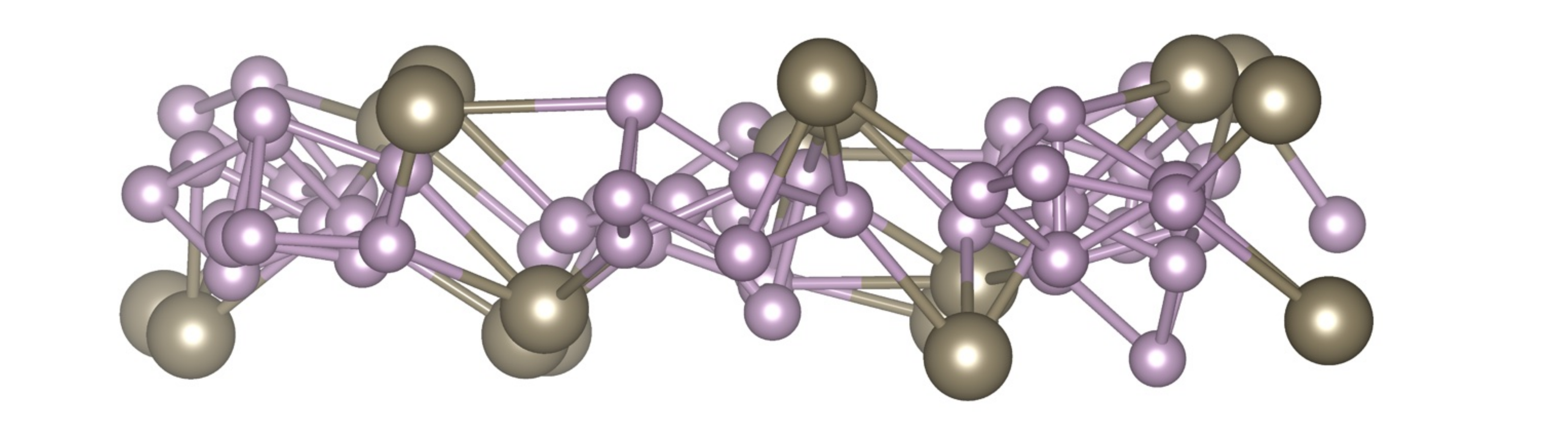}}\\
\sidesubfloat[d]{\includegraphics[width=4cm]{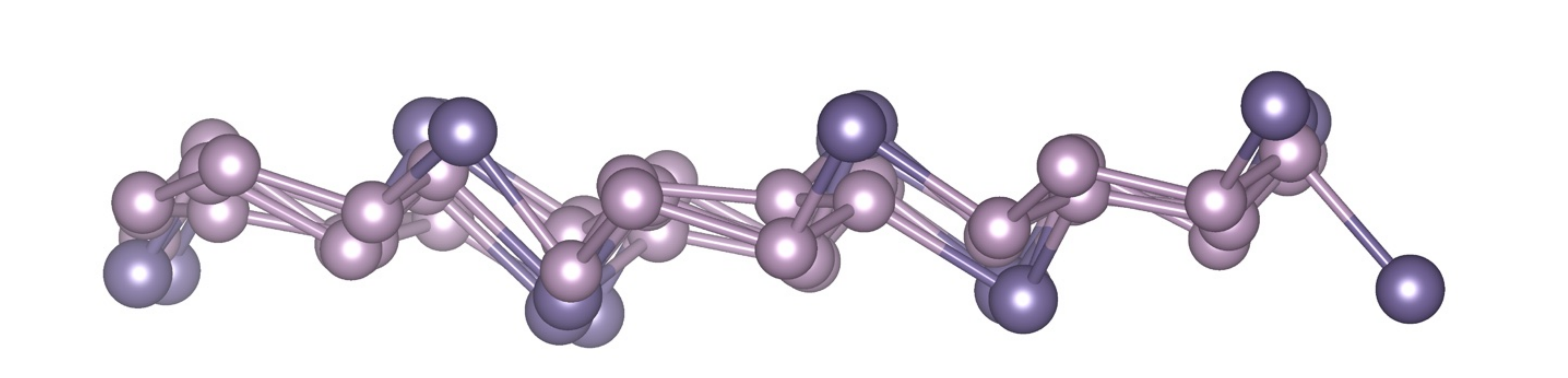}}
\sidesubfloat[e]{\includegraphics[width=4cm]{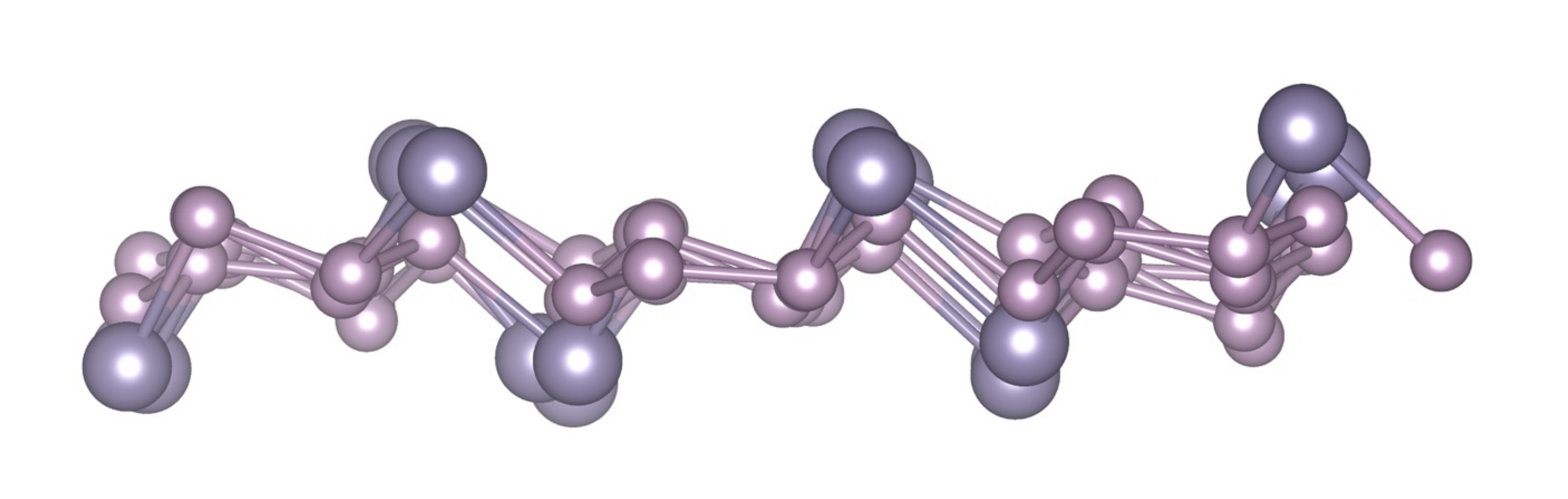}}
\sidesubfloat[f]{\includegraphics[width=4cm]{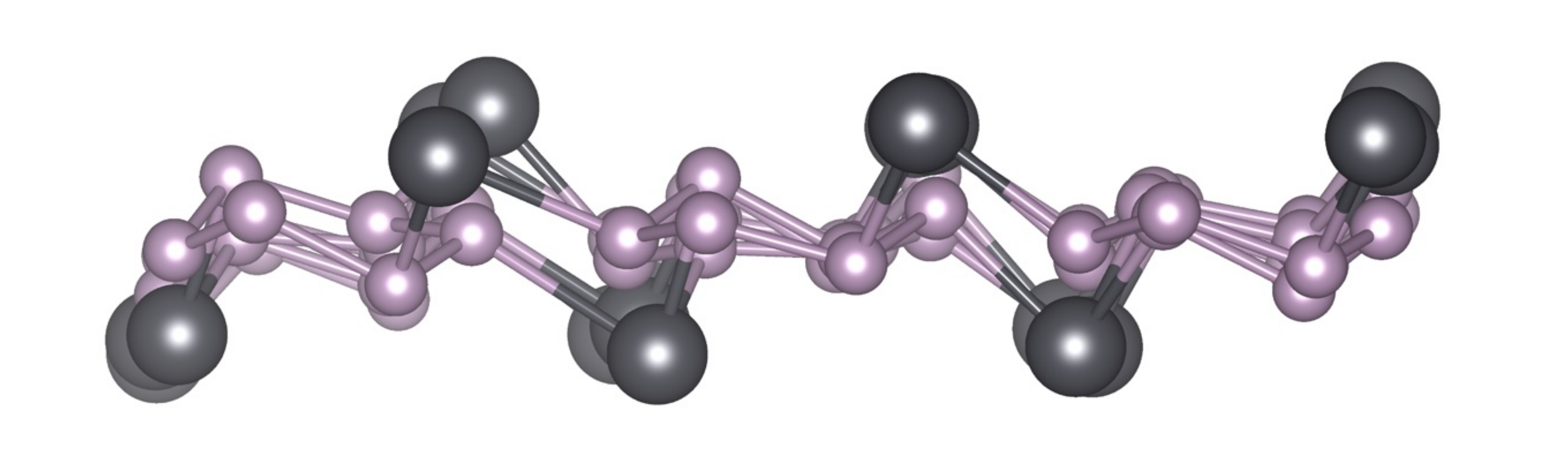}}\\
\sidesubfloat[g]{\includegraphics[width=4cm]{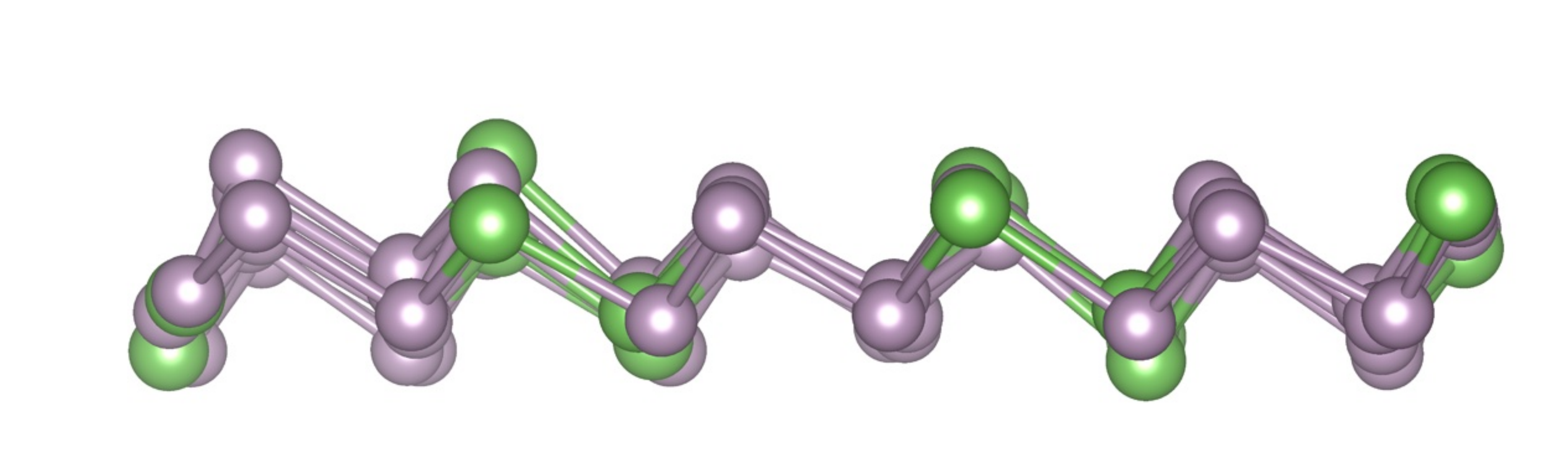}}
\sidesubfloat[h]{\includegraphics[width=4cm]{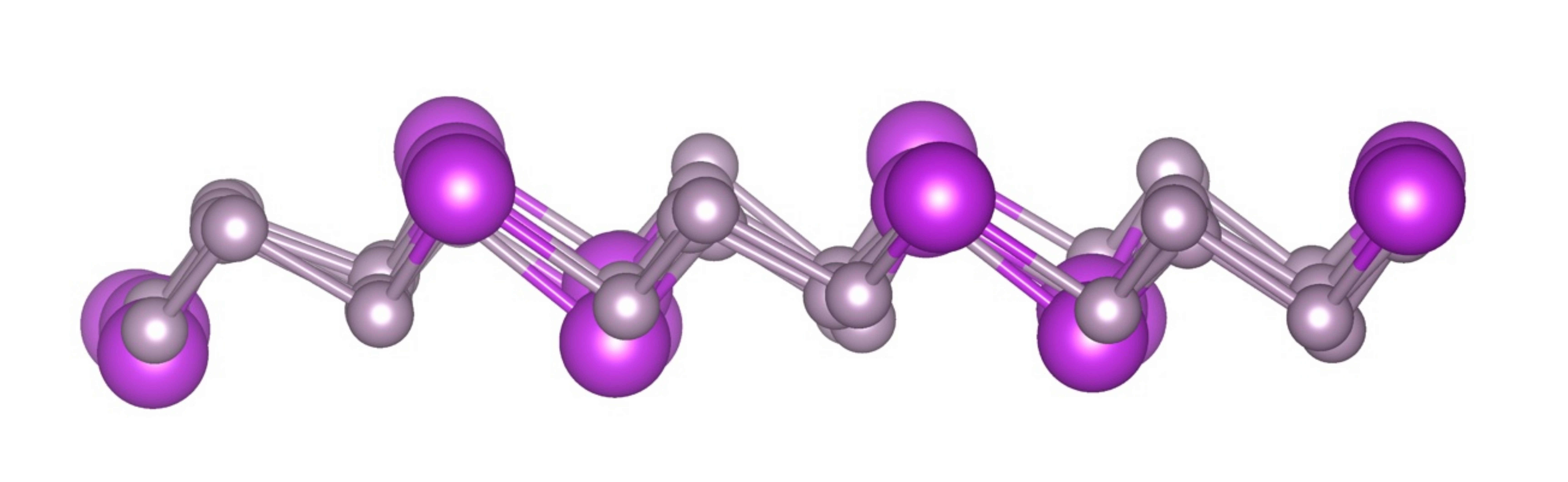}}
\sidesubfloat[i]{\includegraphics[width=4cm]{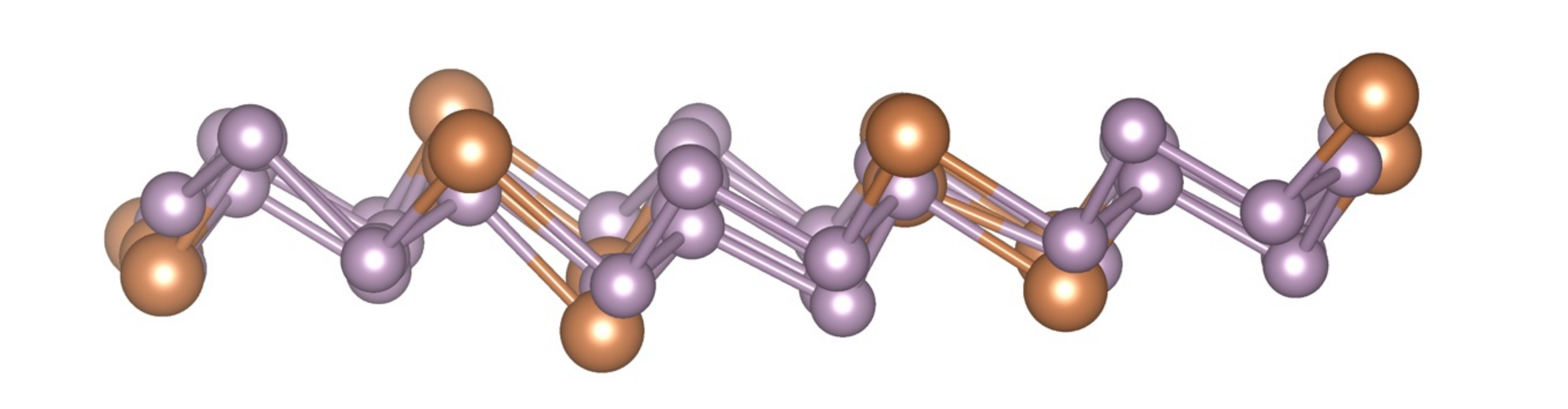}}
\caption{\label{fig:md_structures} AIMD calculations for XP$_3$ (X= Ga, In, Tl, Ge, Sn, Pb, As, Bi, Sb) structures. Snapshots shown at 10\,ps.}
\end{figure}

\begin{figure}[H]
\sidesubfloat[a]{\includegraphics[width=4cm]{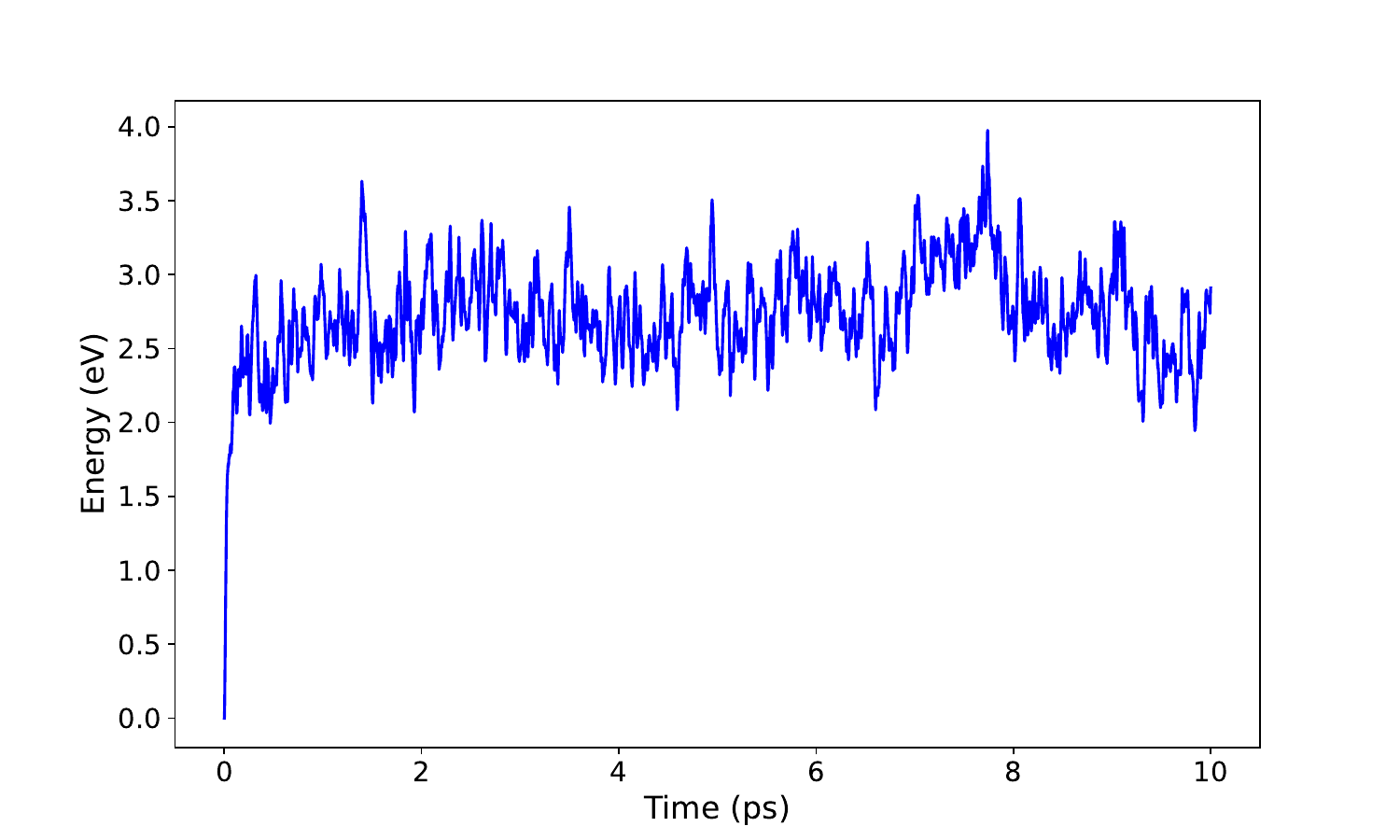}}
\sidesubfloat[b]{\includegraphics[width=4cm]{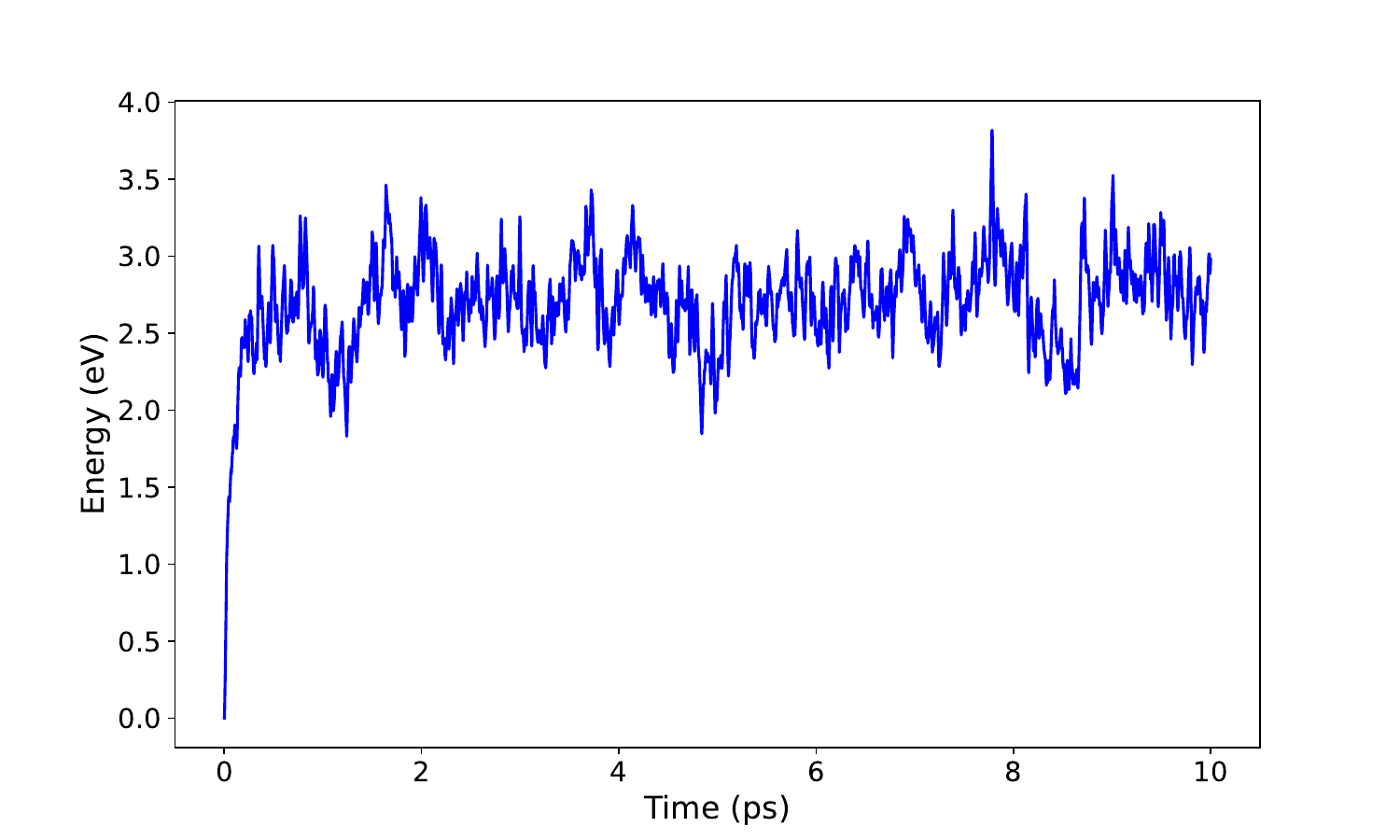}}
\sidesubfloat[c]{\includegraphics[width=4cm]{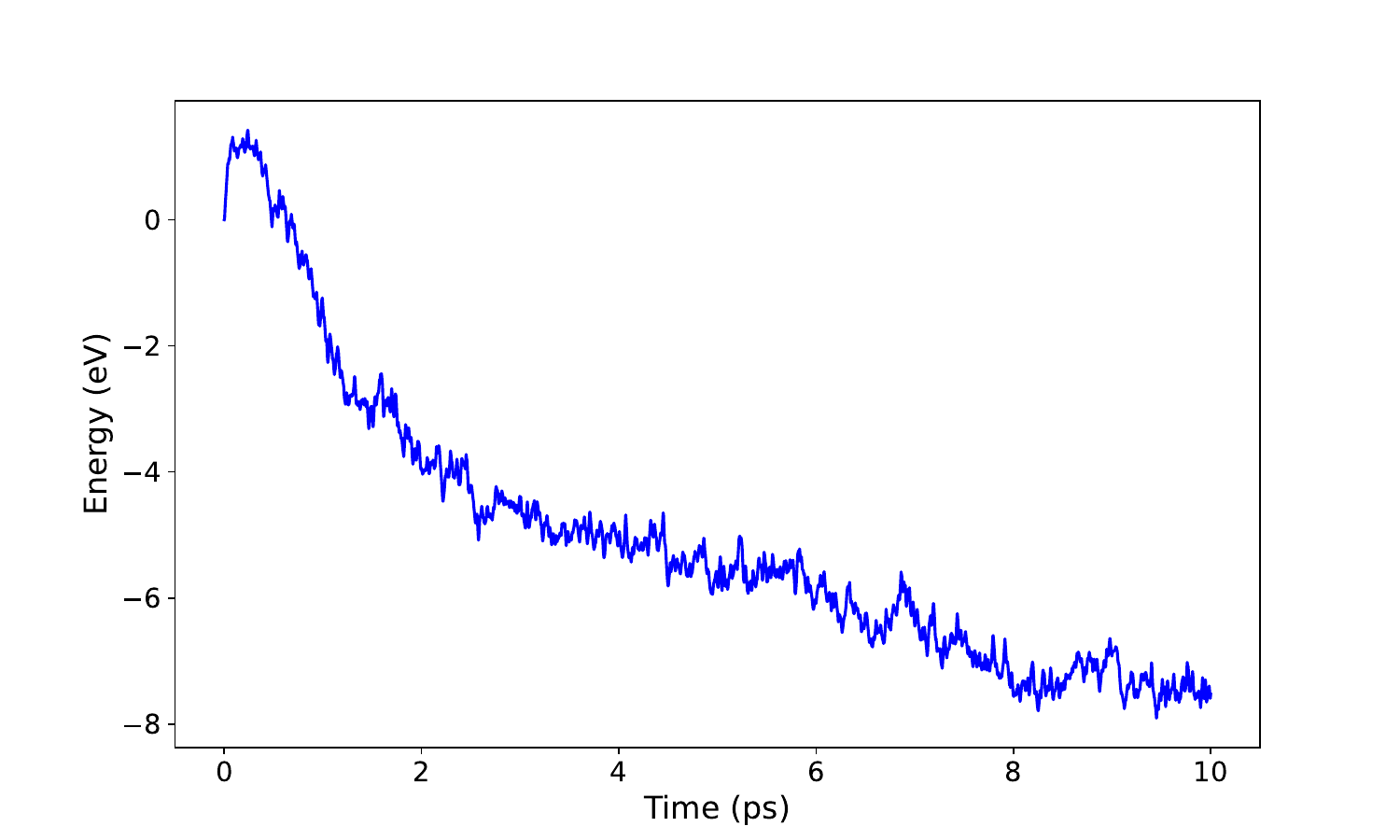}}\\
\sidesubfloat[d]{\includegraphics[width=4cm]{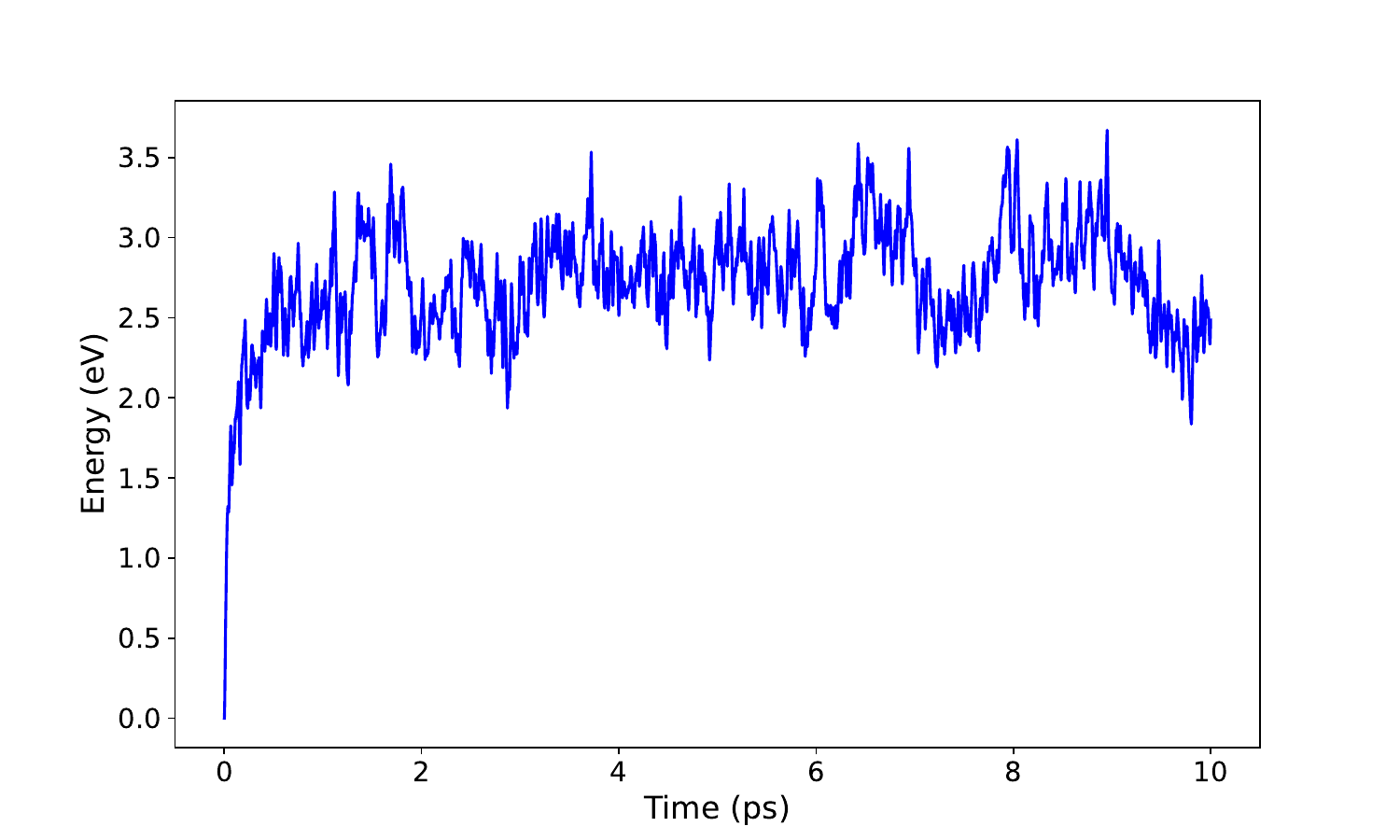}}
\sidesubfloat[e]{\includegraphics[width=4cm]{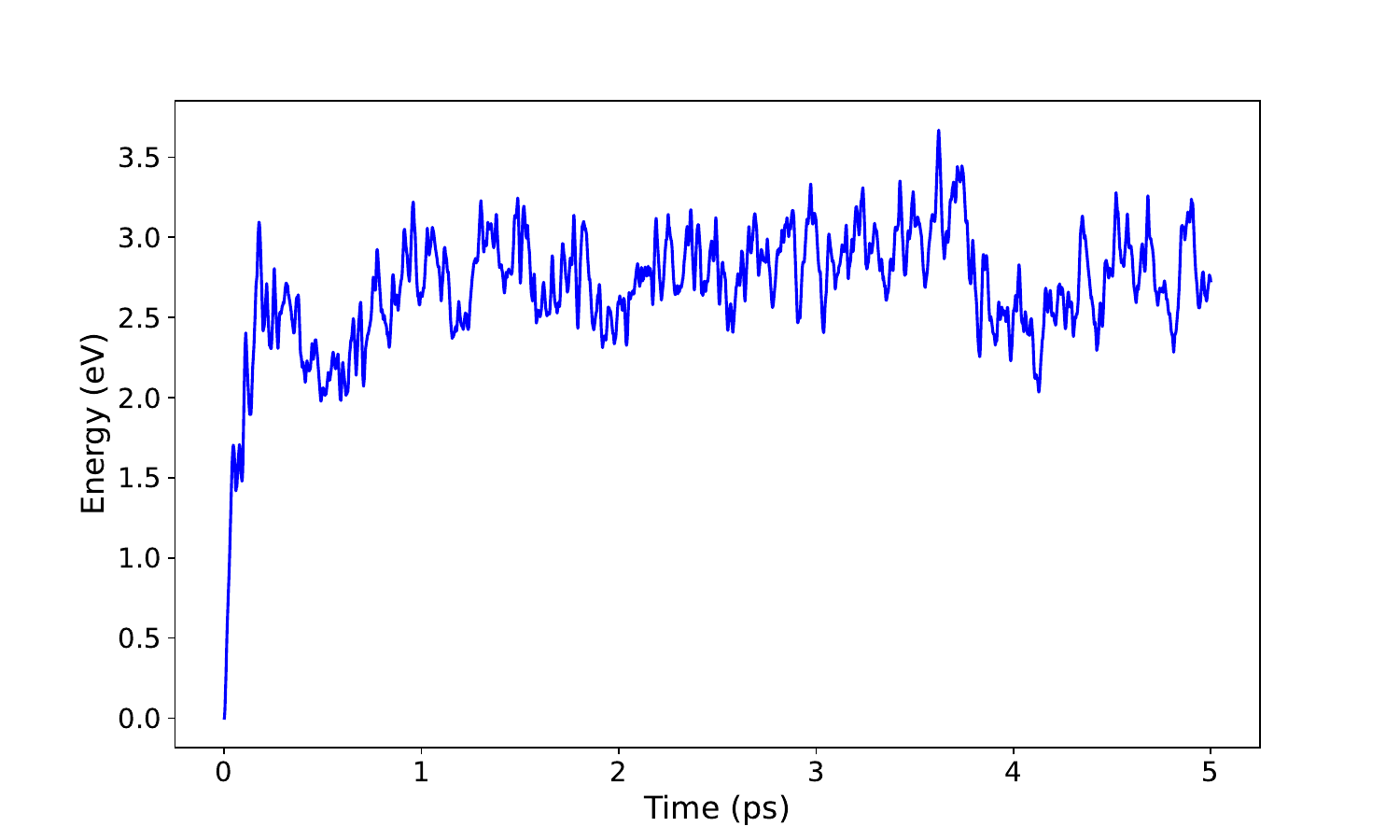}}
\sidesubfloat[f]{\includegraphics[width=4cm]{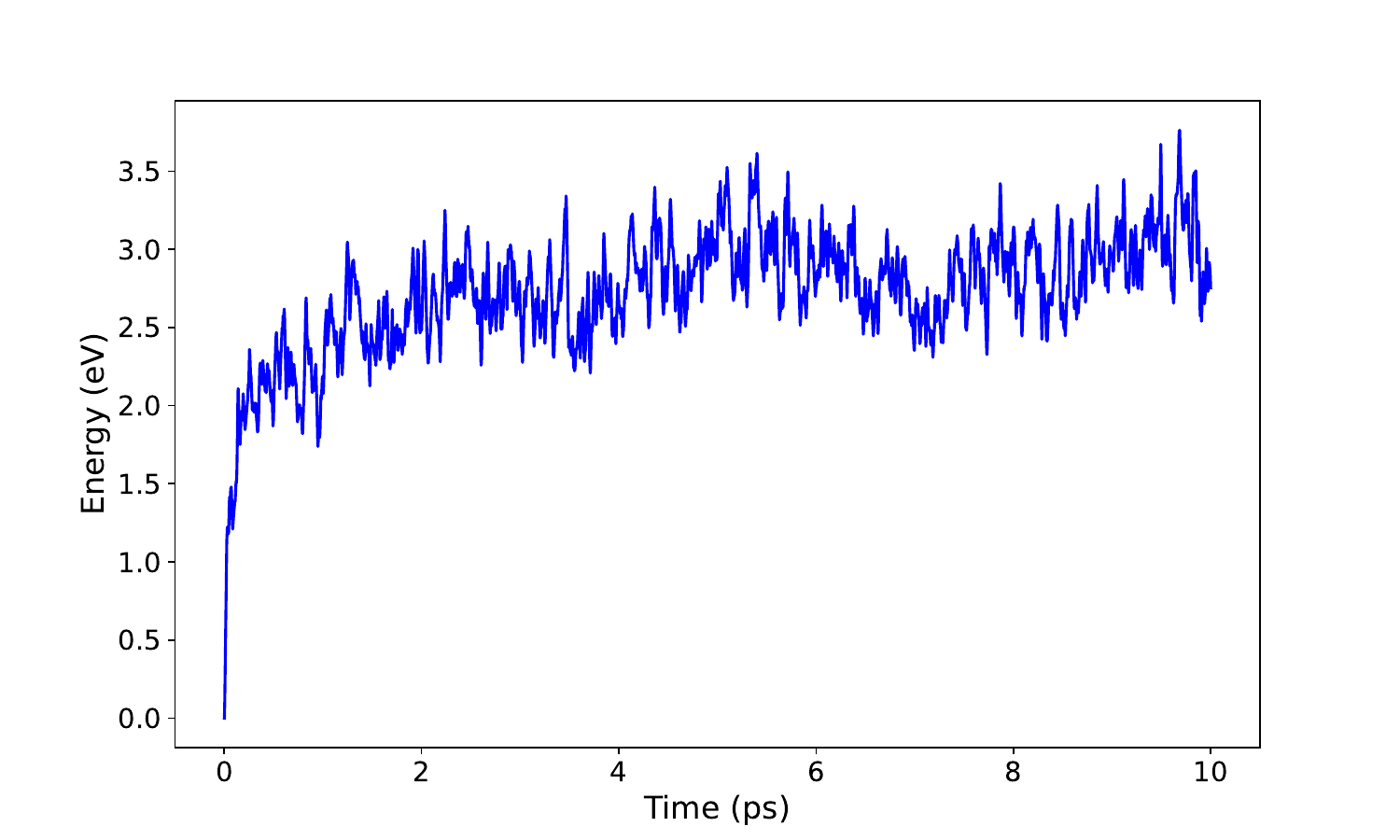}}\\
\sidesubfloat[g]{\includegraphics[width=4cm]{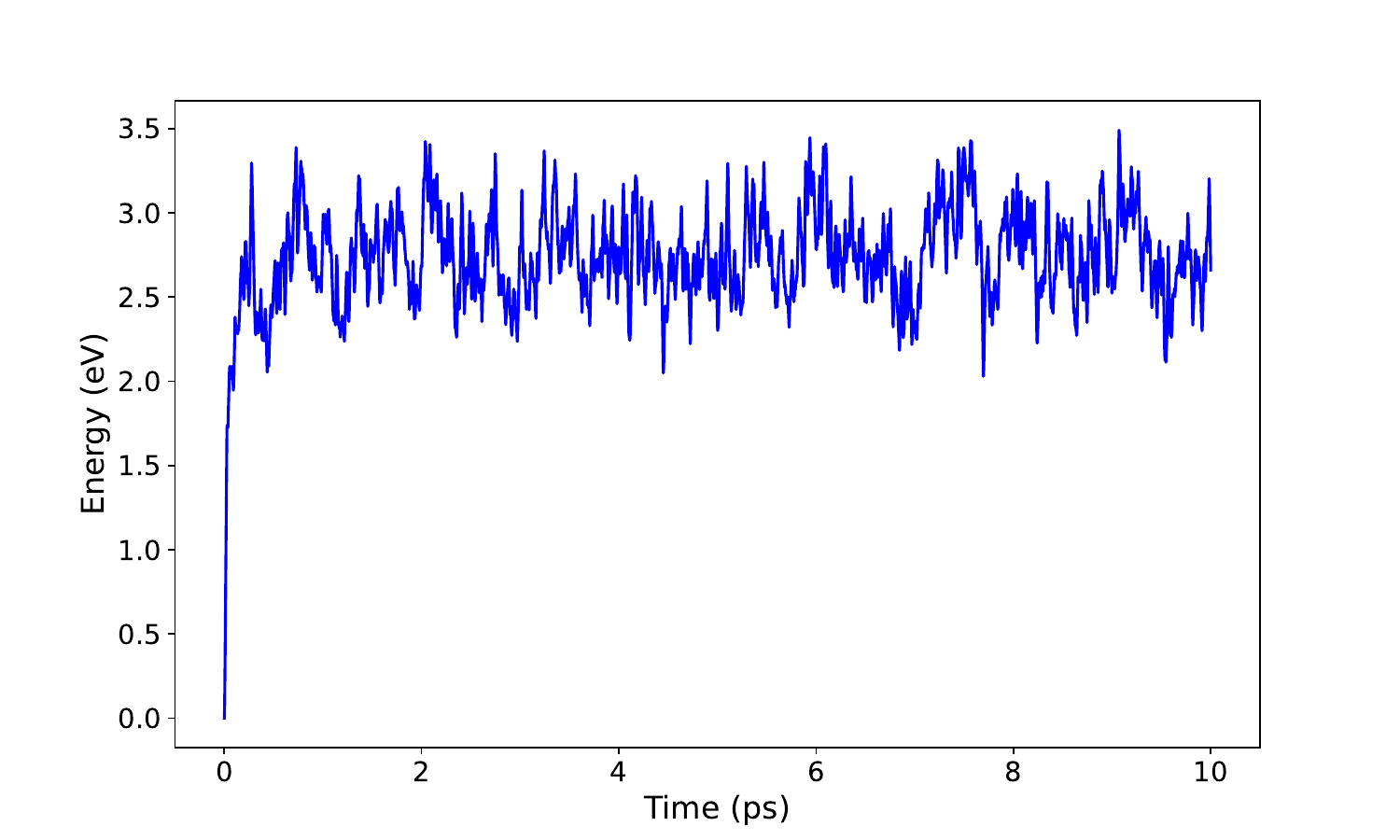}}
\sidesubfloat[h]{\includegraphics[width=4cm]{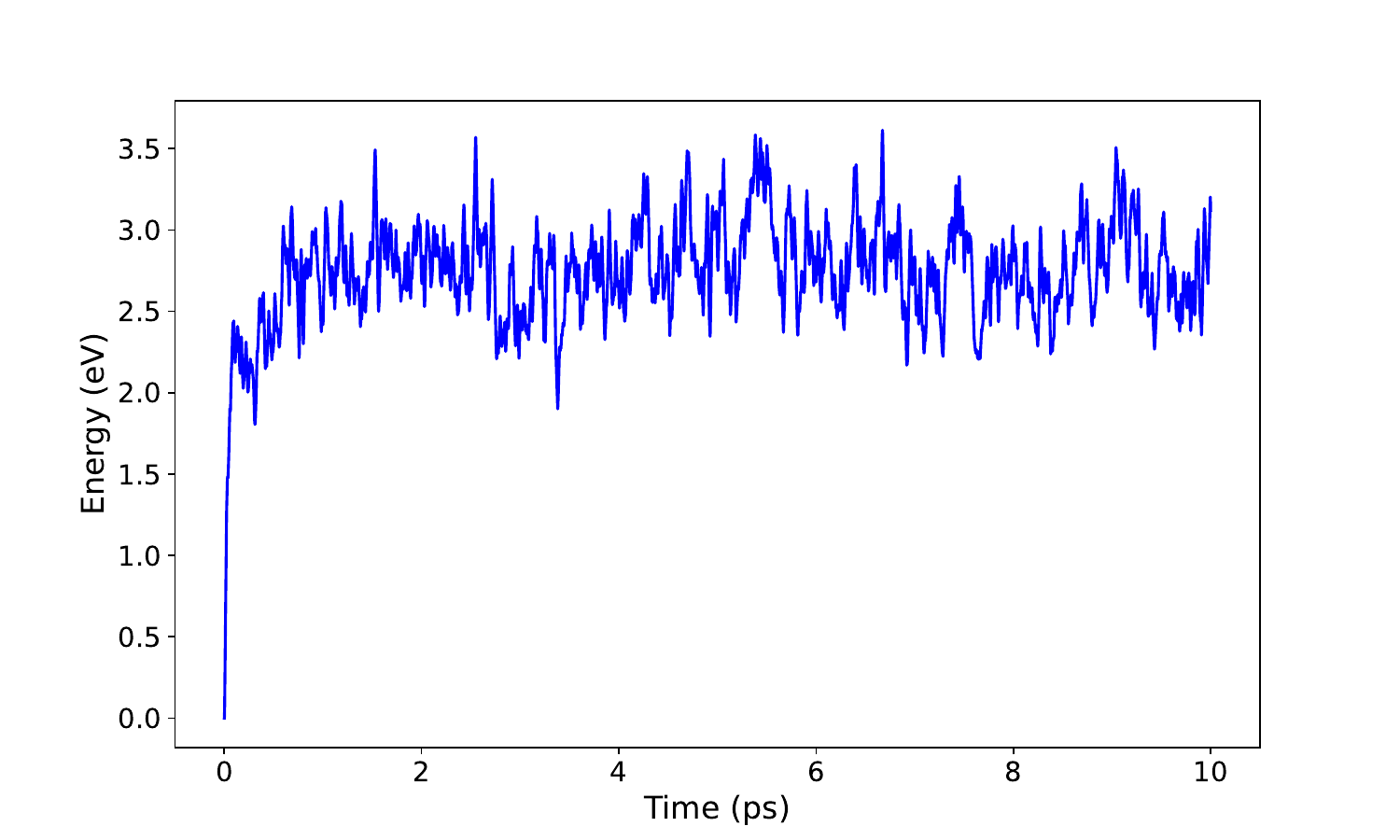}}
\sidesubfloat[i]{\includegraphics[width=4cm]{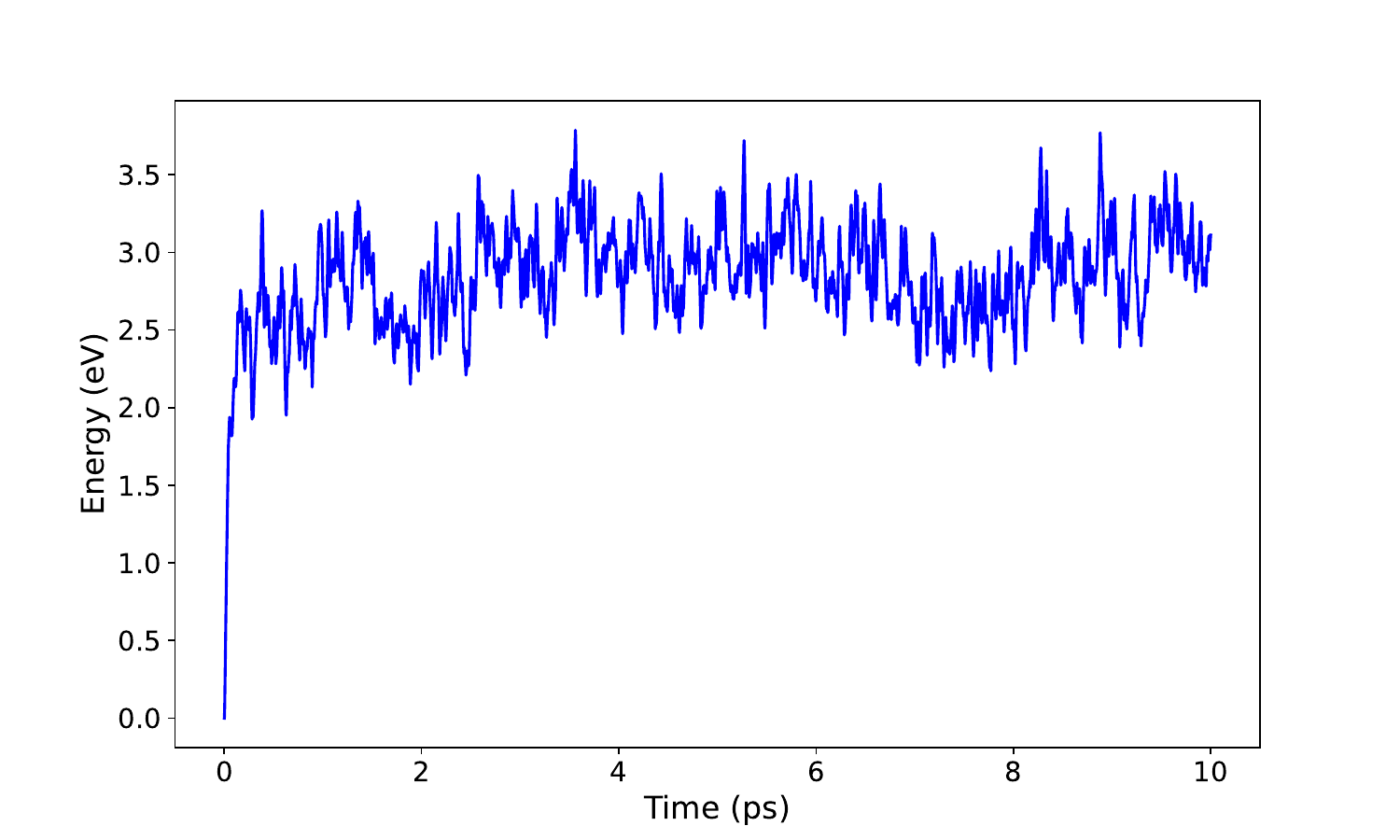}}
\caption{\label{fig:md_energy} Total energy profile of AIMD calculations for XP$_3$  (X= Ga, In, Tl, Ge, Sn, Pb, As, Bi, Sb) monolayers.}
\end{figure}

\begin{figure}[H]
\sidesubfloat[a]{\includegraphics[width=4cm]{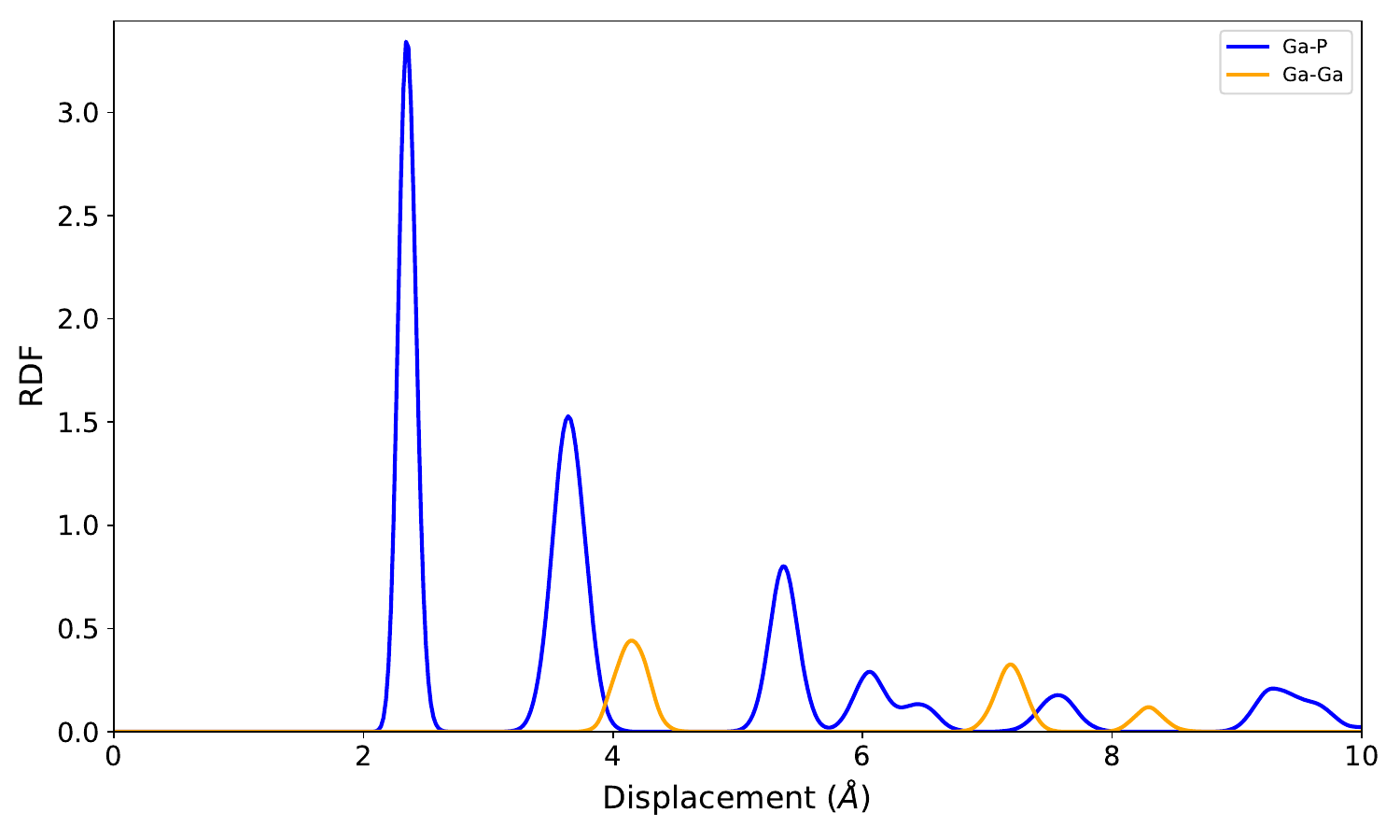}}
\sidesubfloat[b]{\includegraphics[width=4cm]{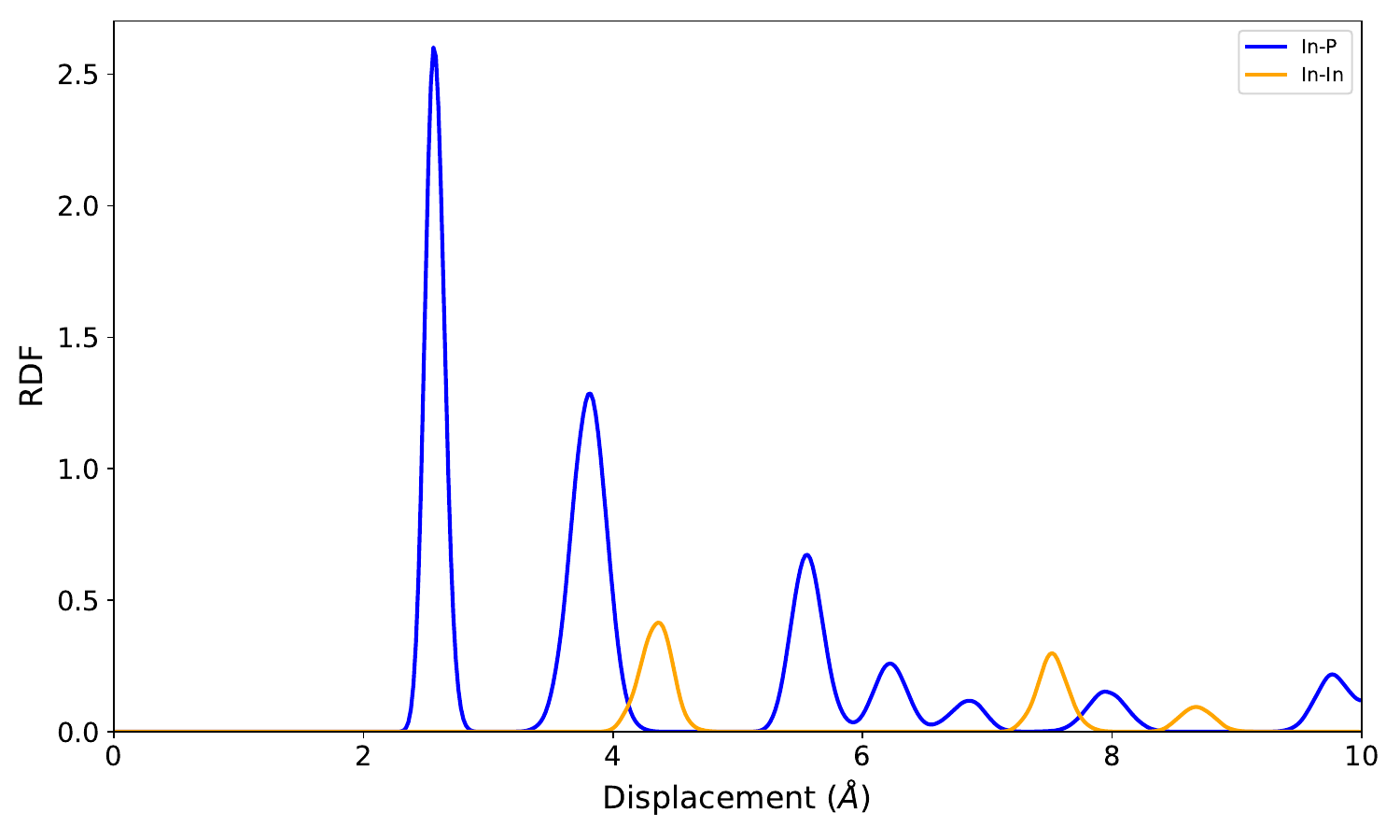}}
\sidesubfloat[c]{\includegraphics[width=4cm]{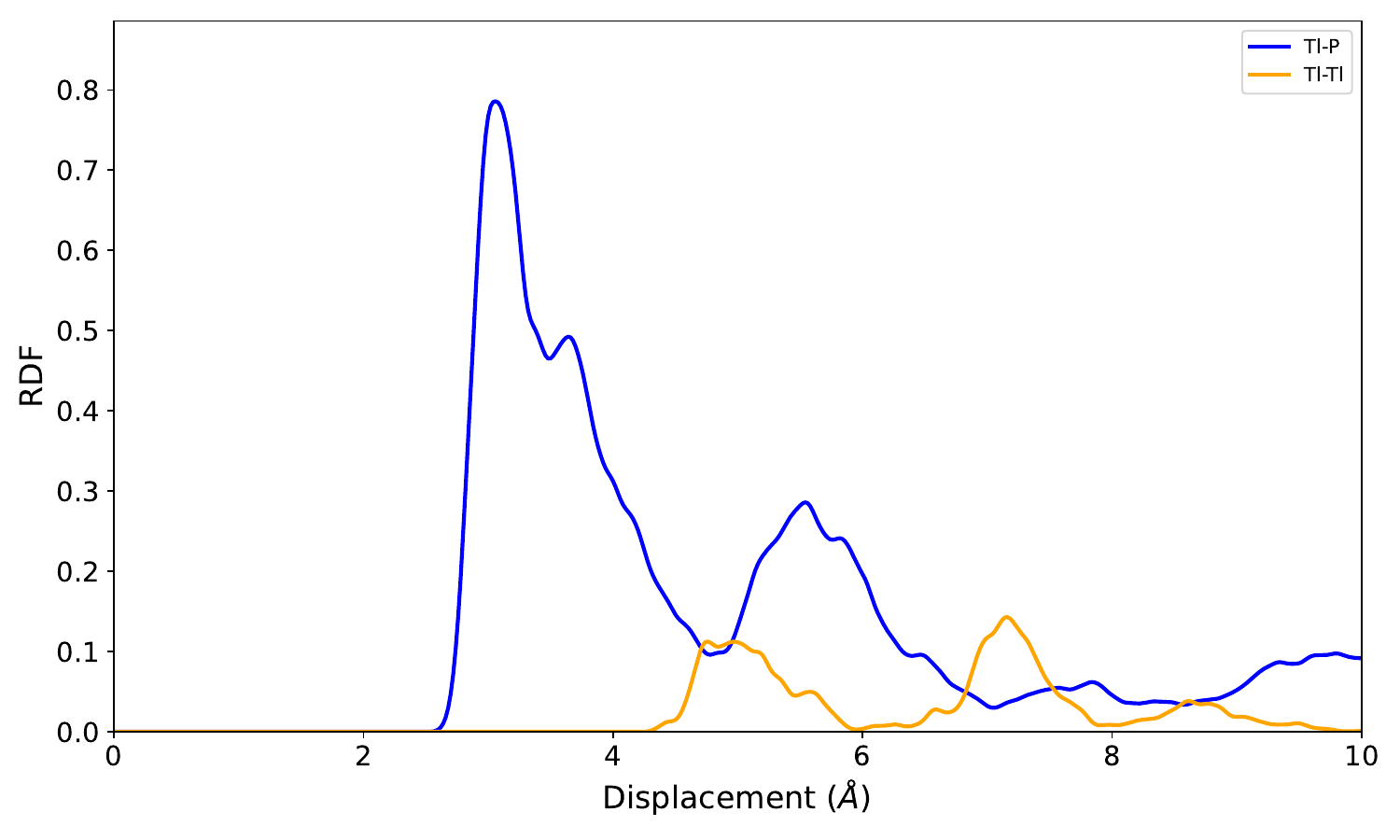}}\\
\sidesubfloat[d]{\includegraphics[width=4cm]{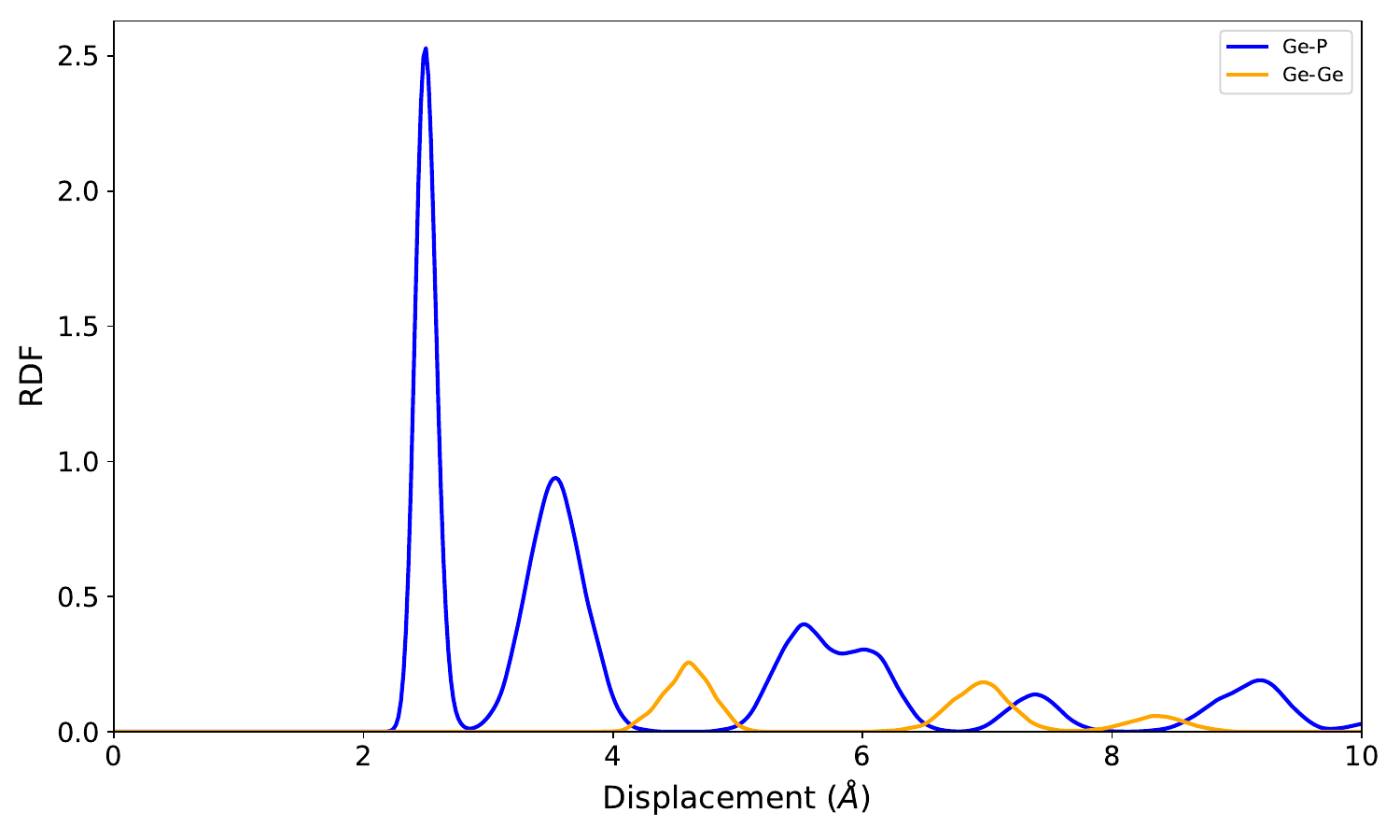}}
\sidesubfloat[e]{\includegraphics[width=4cm]{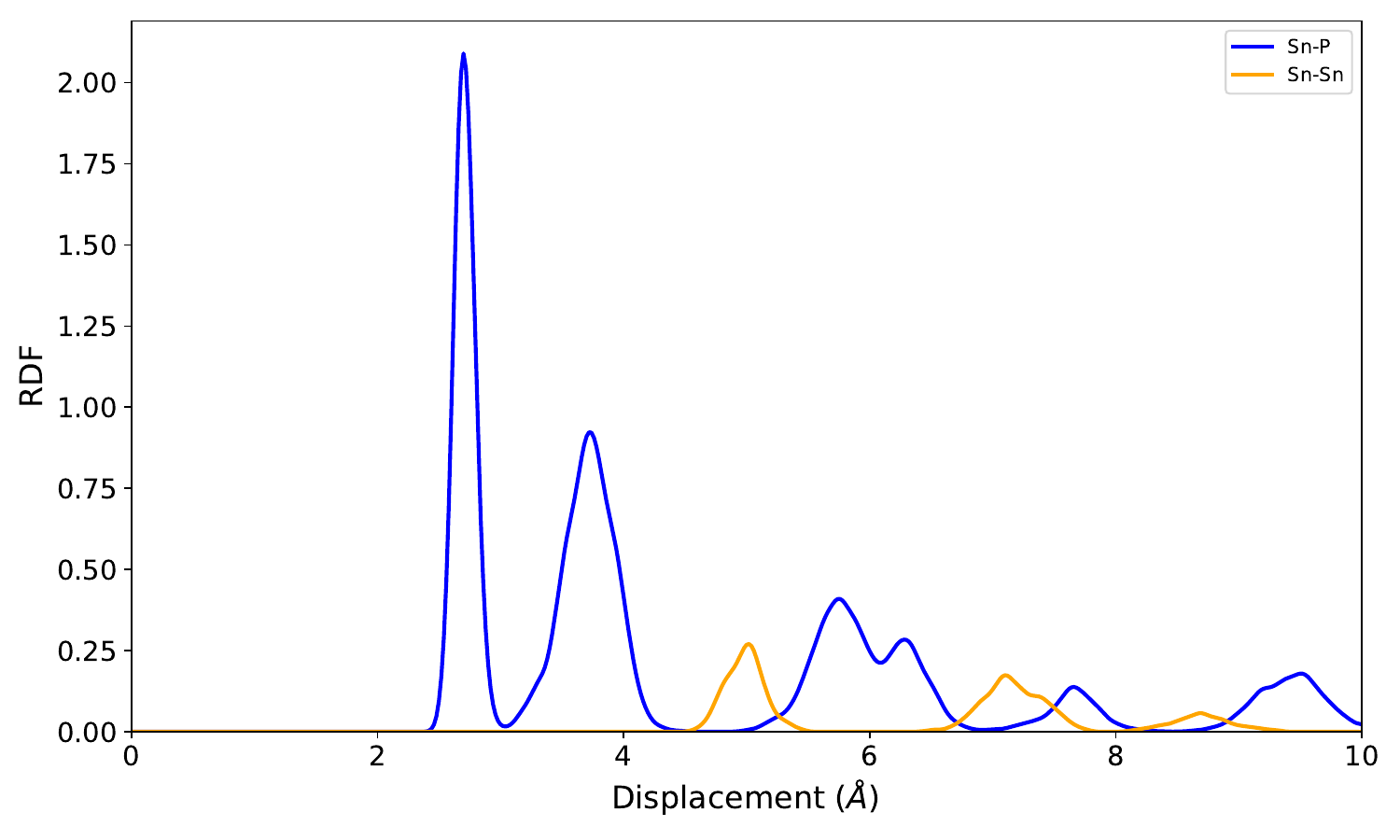}}
\sidesubfloat[f]{\includegraphics[width=4cm]{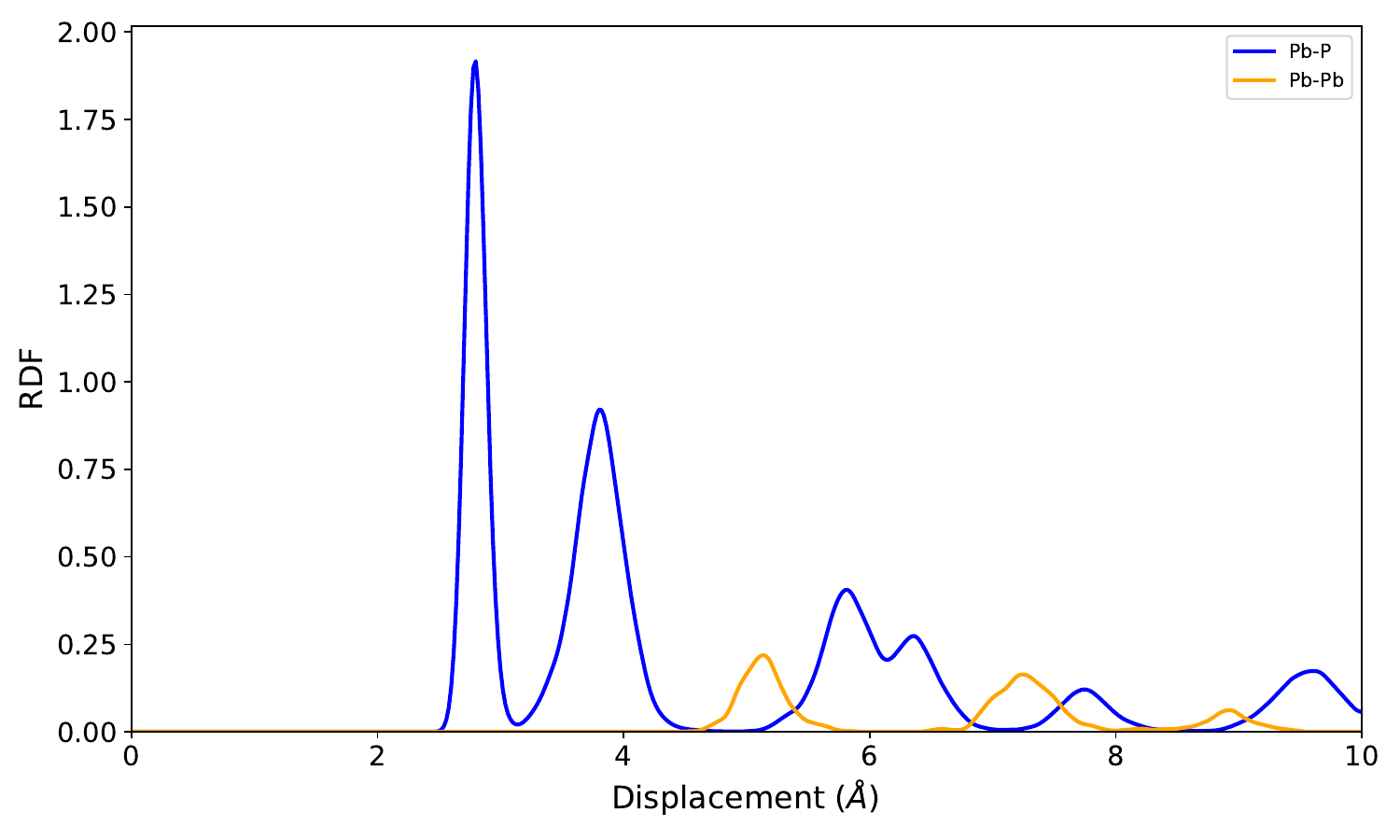}}\\
\sidesubfloat[g]{\includegraphics[width=4cm]{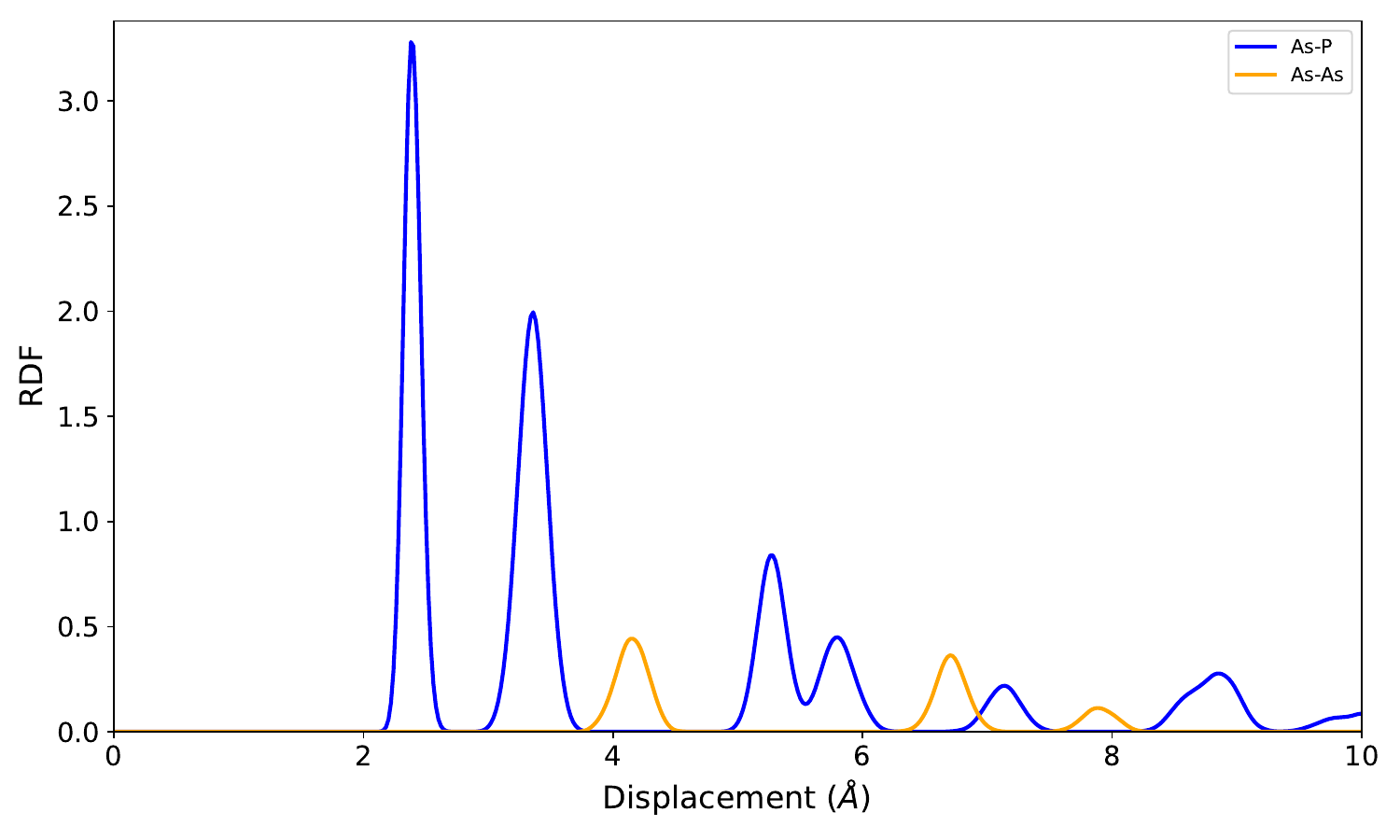}}
\sidesubfloat[h]{\includegraphics[width=4cm]{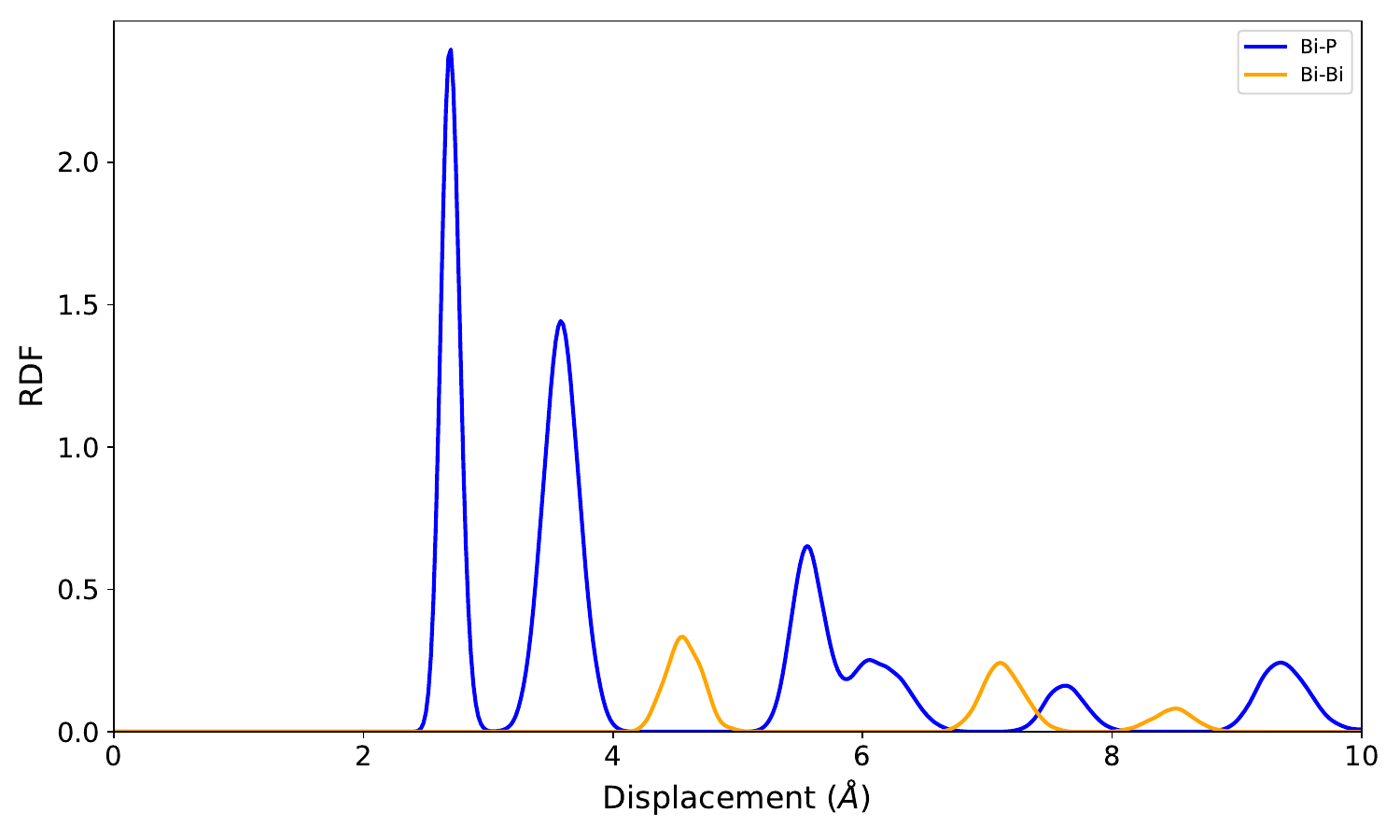}}
\sidesubfloat[i]{\includegraphics[width=4cm]{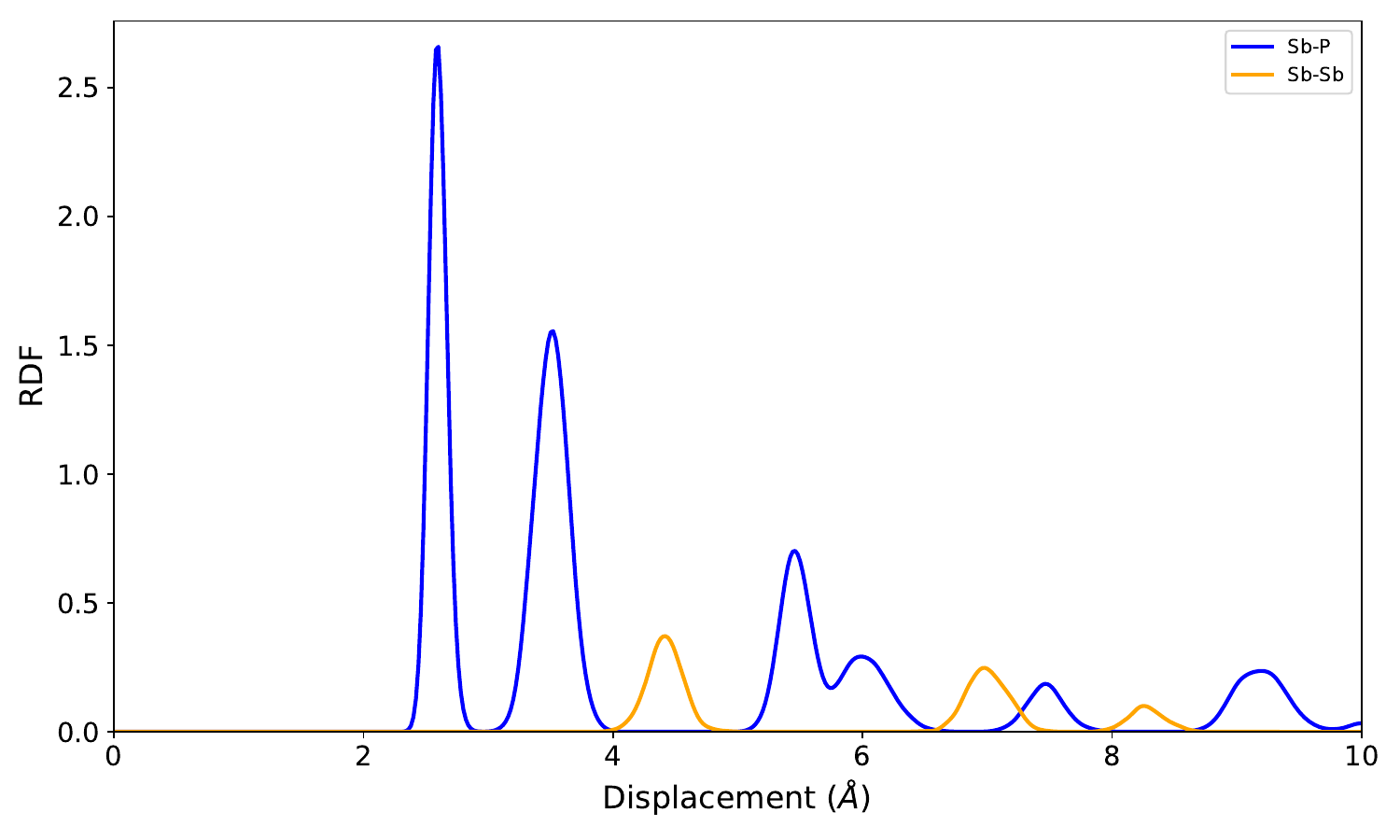}}
\caption{\label{fig:md_rdf} Radial distribution function extracted from the AIMD calculations for XP$_3$  (X= Ga, In, Tl, Ge, Sn, Pb, As, Bi, Sb) monolayers. }
\end{figure}

In order to In Fig.\ref{fig:md_structures} we show the AIMD
calculations for XP$_3$ (X= Ga, In, Tl, Ge, Sn, Pb, As, Bi, Sb)
structures. Snapshots were taken at the final step at 10\,ps. One can
see that all structures maintain the hexagonal symmetry. However, the
bonds are somewhat distorted from the pristine structure, indicating
the metastability of XP$_3$ monolayers at room temperature. In
Fig.\ref{fig:md_energy} we show the total energy profile of the AIMD
calculations for XP$_3$ monolayers for 10\,ps simulation time. In
Fig.\ref{fig:md_rdf} we show the total energy profile of the AIMD
calculations for XP$_3$ monolayers for 10\,ps simulation time.
Surprisingly the TlP$_3$ monolayer shows a strong distortion, apparently yielding to a phase separation of the P and Tl layers. Further
investigations need to be performed to understand why the phase
segregation occurs.

\section{Excitonic and Optical Properties}

The excitonic effects are fundamental for an accurate description of the linear optical response in 2D materials.\cite{Haastrup_042002_2018,Dias_3265_2021} Due the significant quantum confinement in their non-periodic direction, these quasi-particle effects can results in a significant optical band gap red-shift around \SIrange{100}{500}{\milli\electronvolt}.\cite{Dias_3265_2021,Moujaes_111573_2023,Dias_8572_2024}

\begin{table}[H]
\centering
\caption{Excitonic properties obtained using MLWF-TB+BSE at DFT-HSE06: fundamental band gap $E_{g}$, direct band gap $E^{d}_{g}$, exciton ground state $Ex_{gs}$ and direct exciton ground state $Ex^{d}_{gs}$. The exciton binding energy $Ex_{b}$ is  calculated as $E_{g}-Ex_{gs}$. All calculations include SOC.}
\begin{tabular*}{12cm}{@{\extracolsep{\fill}}l cccccc}\toprule
&      & $E_{g}$ (eV) & $E^{d}_{g}$ (eV) & $Ex_{gs}$ (eV) & $Ex^{d}_{gs}$ (eV)  & $Ex_{b}$ (meV) \\ \toprule
          &\ce{GaP3} &1.44  &1.61  &1.10  &1.47  &334.91 \\ 
Group III &\ce{InP3} &1.31  &1.31  &1.16  &1.27  &148.84 \\
          &\ce{TlP3} &0.89  &0.95  &0.56  &0.56  &334.43 \\ \midrule
          &\ce{GeP3} &0.53  &1.12  &0.18  &0.65  &347.43 \\ 
Group IV  &\ce{SnP3} &0.70  &1.38  &0.34  &0.87  &358.22 \\
          &\ce{PbP3} &0.83  &1.43  &0.50  &0.98  &328.93 \\ \midrule
          &\ce{AsP3} &2.59  &2.72  &2.37  &2.49  &211.84 \\ 
Group V   &\ce{BiP3} &1.97  &2.23  &1.85  &1.89  &123.42 \\
          &\ce{SbP3} &2.31  &2.46  &2.07  &2.14  &242.60 \\ 
 \bottomrule
\end{tabular*}
\label{tab:exciton_data}
\end{table}

\begin{figure}[H]
\includegraphics[width=1.0\linewidth]{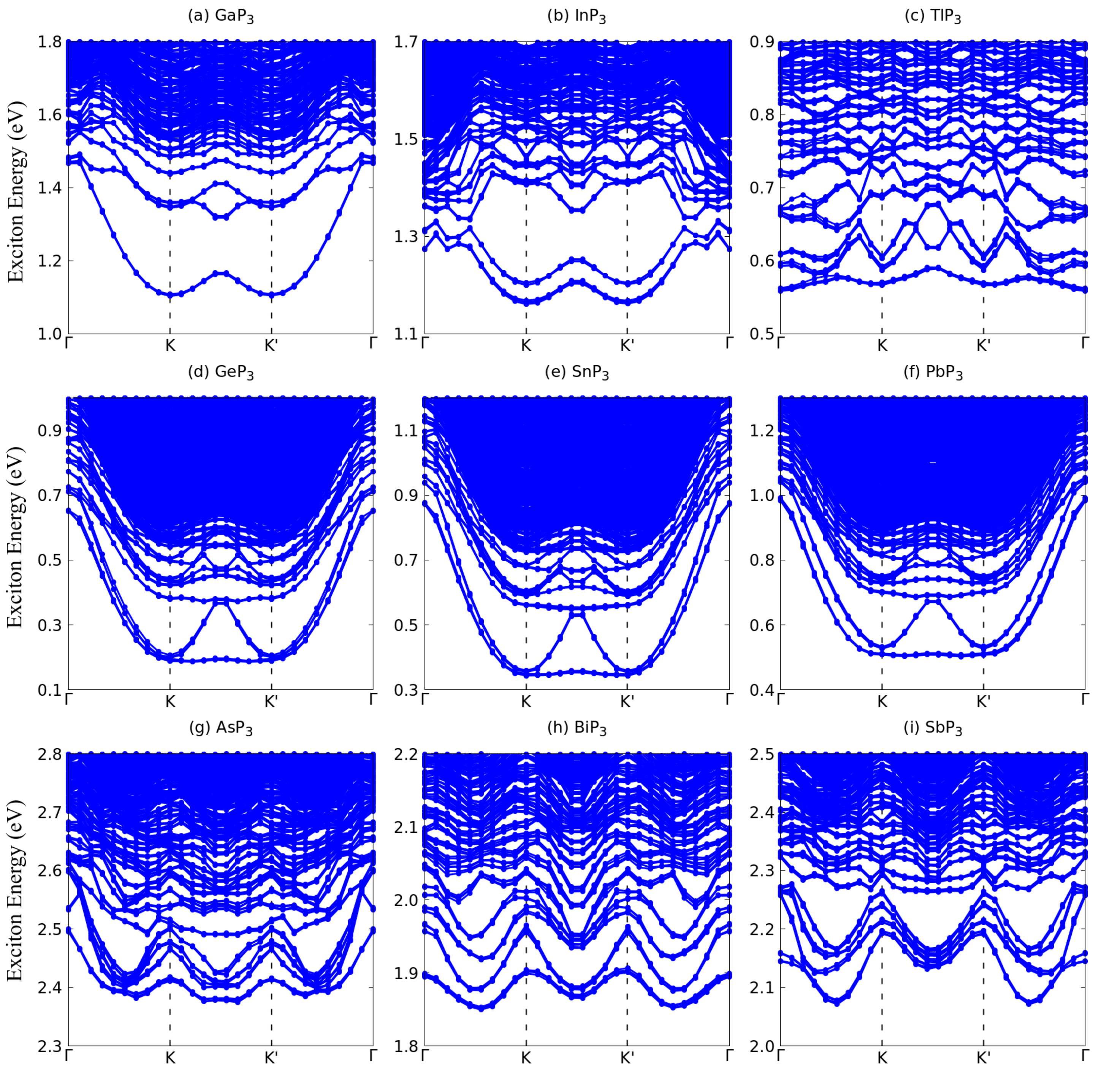}
\caption{\label{fig:exc_bands} Exciton band structure of a) \ce{GaP$_3$}, (b) \ce{InP$_3$}, (c) \ce{TlP$_3$} d) \ce{GeP$_3$}, (e) \ce{SnP$_3$} and (f) \ce{PbP$_3$} (g) \ce{AsP$_3$}, (h) \ce{BiP$_3$} and (i) \ce{SbP$_3$}. All calculations were carried out using MLWF-TB+BSE at DFT-HSE06 and include SOC.}
\end{figure}

We start our investigation of the excitonic effects in \ce{XP3}
monolayers from the exciton band structure, shown in
Fig.~\ref{fig:exc_bands}, the exciton binding energy ($Ex_{b}$) was
estimated by the difference of the fundamental band gap ($E_{g}$) and
the exciton ground state ($Ex_{gs}$) energies, shown in
Table~\ref{tab:exciton_data}. As expected, for all monolayers, except
\ce{TlP3} the exciton ground state was indirect (i.e out of $\Gamma$
point), which can be justified from the indirect electronic band gap
of these system, being, in this case, \ce{InP3} the only exception, as
it is a direct band gap semiconductor. This kind of behavior for
\ce{TlP3} and \ce{InP3} was not common, but similar cases was
previously reported in the
literature,\cite{Lu_5204_2022,Santos_074301_2023,Santos_3677_2023}
being justified by the different Coulomb strength in each
electron-hole pair and its correspondent wave-functions. In group III
for \ce{GaP3} and \ce{TlP3} the exciton binding energy lies around
\SI{330}{\milli\electronvolt}, a behavior that is different from
\ce{InP3} that has \SI{148.84}{\milli\electronvolt}, a difference that
could be justified by its direct electronic band gap nature and the
electrons and holes wavefunctions. In group IV for the 3 monolayers
$Ex_{b}$ lies in the range of
\SIrange{328.93}{358.22}{\milli\electronvolt} and in Group V
\SIrange{123.42}{242.60}{\milli\electronvolt}, showing that the
exciton binding energy is high dependent of \ce{XP3} chemical
composition. These presence of indirect excitonic ground state, also
indicates the possibility of phonon-assisted optical transitions,
which absorption peaks with excitation energies lowers than the ones
predicted by the optical band gap (estimated by the direct excitonic
ground state $Ex_{gs}^{d}$) at BSE level.

Complementary we can also see that for Group III and IV, the
exciton ground state was degenerated and localized in the vicinity of
$K$ and $K'$ high symmetry points, except for \ce{TlP3} which was
located at $\Gamma$. For Group V the behavior is different, with a
degenerated excitonic ground state located between $\Gamma$ and $K/K'$
high symmetry points, except for \ce{AsP3} where it is non-degenerated
and located between $K$ and $K'$ points. This degeneracy for the
excitonic states at $K/K'$ can be justified by the time-reversal
symmetry in these monolayers.

\begin{figure}[H]
\includegraphics[width=1.0\linewidth]{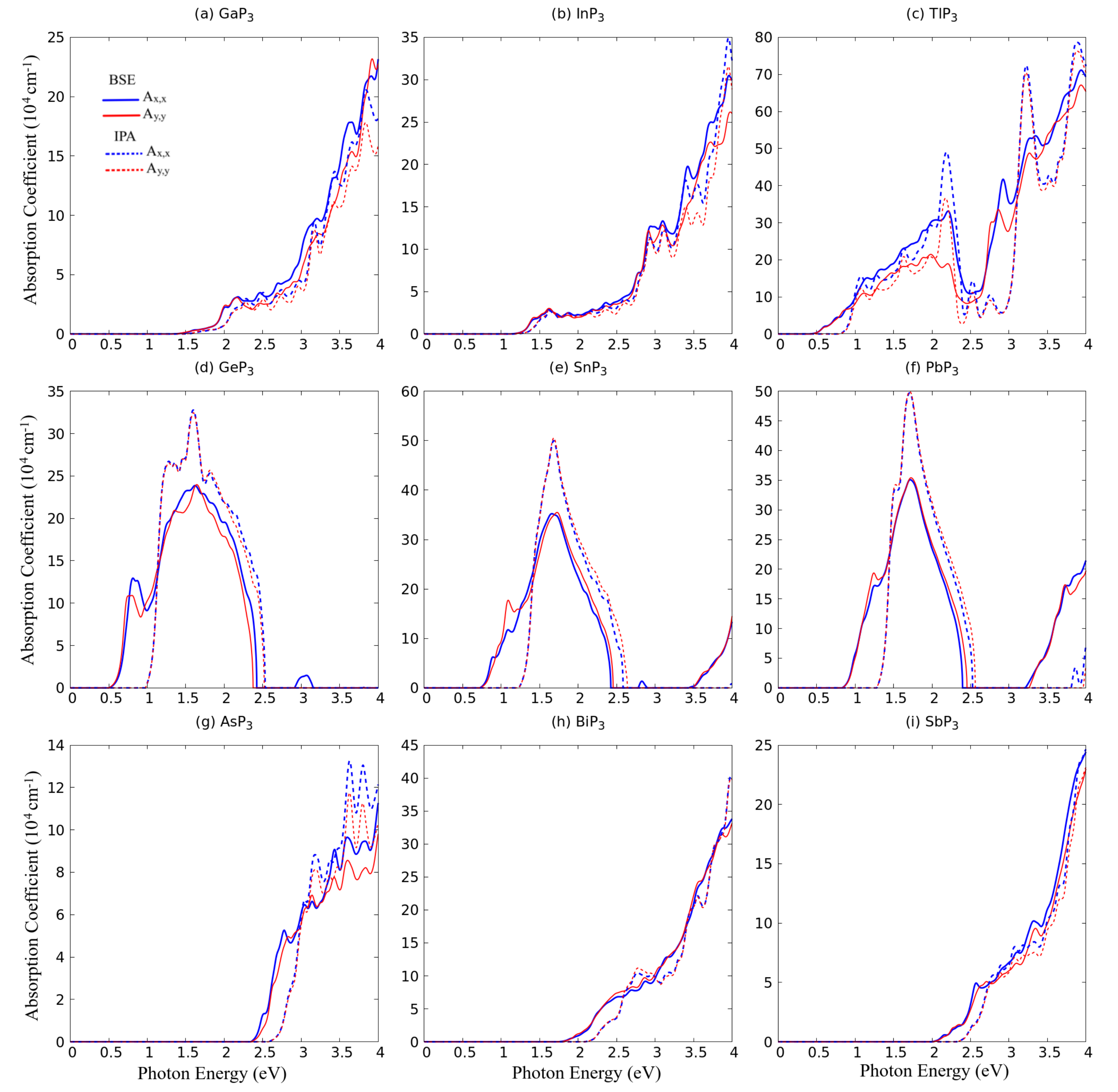}
\caption{\label{fig:abs_coef} Absorption coefficients for XP$_3$ compounds at IPA (dashed curves) and BSE (solid curves). The results for response to a linear polarized light along $\hat{x}$ (blue curves) and $\hat{y}$ (red curves) directions are shown.}
\end{figure}

The linear optical response was shown through the absorption coefficient, considering an incident polarized light at $\hat{x}$ (blue curves) and $\hat{y}$ (red curves) directions, depicted in Fig.~\ref{fig:abs_coef} at IPA (dashed curves) and BSE (solid curves) levels, these results are complemented by the refractive index and reflectibility in SI section S4. The absorption spectrum also shows that Group III and V  absorbs in the visible and UV regions, except for \ce{TlP3} that also absorbs in the infrared region; Group V in infrared and visible regions, with a weak optical response in UV region. At IPA level group IV and group V, except from \ce{AsP3} are isotropic, in group III all monolayers are anisotropic. When quasi-particle effects are considered a small optical anisotropy emerges in the isotropic systems, despite the linear response at  $\hat{x}$ and  $\hat{y}$ are not so much different. For all systems we can easily observes the red-shift in the optical band gap due excitonic effects, despite the optical anisotropy, the optical band gap doesn't changes due incident light polarization, excitonic effects also lowers the absorption coefficient for higher photon excitations, an effect that could be expected due the optical band gap red-shift.

\section{Insights into Solar Harvesting Efficiency}

The \ce{XP3} monolayer solar harvesting efficiency was estimated by the PCE, calculated at IPA and BSE levels, using the SQ-limit \cite{Shockley_510_1961} and SLME \cite{Yu_068701_2012} methods, as shown in Table~\ref{table:pce_ipa_bse}. In SQ-limit the PCE was estimate straightforwardly from the optical band gap from IPA and BSE calculations, in SLME approach beyond the optical band gap, we also need additional factors such as layer thickness, the nature of the fundamental band gap (IPA) or excitonic ground state (BSE) and the monolayer total absorption spectrum. It's also important to understand that the present results, corresponds to the superior limit of the solar harvesting efficiency of these materials, and other engineering problems in the solar device can lowers these values. Achieving values closer to the shown upper limit can often requires years of research, being one example the case of \ce{MaPbI3} perovskite which has a significant increase of experimental PCE in the last decade being from \SIrange{3.8}{25.2}{\percent}.\cite{Kojima_6050_2009, Frolova_6772_2020} 

\begin{table}[H]
\centering
\caption{Maximum achieved PCE at the IPA and BSE levels,  PCE$^{\text{SLME}}$ (\si{\percent}), power conversion efficiency determined by SLME and considering that \SI{100}{\percent} of photon absorbance starts from direct band gap, PCE$_{max}^{\text{SLME}}$ (\si{\percent}) obtained in the Shockley--Queisser limit (considering direct band gap), and PCE$^{\text{SQ}}$ (\si{\percent}), all calculated at $T = \SI{300}{\kelvin}$.}
\begin{tabular}{lcccccccccc} \toprule
& &  & IPA & &&  & BSE & \\ \cmidrule{3-5} \cmidrule{7-9}
& System & PCE$^{\text{SLME}}$ &PCE$_{max}^{\text{SLME}}$ &PCE$^{\text{SQ}}$  && PCE$^{\text{SLME}}$ &PCE$_{max}^{\text{SLME}}$ &PCE$^{\text{SQ}}$ \\ \midrule
         &\ce{GaP3} &0.10  &25.24 &29.53 & &0.09 &20.99 &31.65 \\
Group III&\ce{InP3} &0.18  &32.23 &32.23 & &0.18 &27.92 &31.88 \\
         &\ce{TlP3} &1.03  &26.81 &29.68 & &0.64 &15.70 &15.70 \\ \midrule
         &\ce{GeP3} &0.36  &7.87 &32.18 & &0.00 &0.05 &19.57 \\
Group IV &\ce{SnP3} &0.70  &10.75 &32.62 & &0.14 &2.17 &26.64 \\
         &\ce{PbP3} &0.86  &13.83 &32.09 & &0.40 &8.02 &30.03 \\ \midrule
         &\ce{AsP3} &0.07  &7.70 &8.22 & &0.10 &11.71 &12.45 \\
Group V  &\ce{BiP3} &0.22  &14.99 &17.58 & &0.29 &23.86 &24.62 \\
         &\ce{SbP3} &0.12 &12.04 &13.03 & &0.15 &18.54 &19.38\\ \bottomrule
\end{tabular}
\label{table:pce_ipa_bse}
\end{table}

At SQ-Limit the PCE$^{\text{SQ}}$ lies between
\SIrange{29}{32}{\percent} at IPA and \SIrange{15}{31}{\percent} at
BSE levels for Group III and IV, in Group V these values are lower,
lying between \SIrange{7}{14}{\percent} at IPA and
\SIrange{12}{19}{\percent} at BSE levels. However these values can be
not realistic for Group IV, as SQ-limit assumes that all photons with
excitation energies higher or equals the optical band gap are
absorbed, but when we look the optical response of this group, we see
that it has regions of the visible and UV spectrum that does not show
optical absorption.

SLME, as considers the optical absorption spectrum, can provide a more
realistic picture of the solar harvesting efficiency of these
monolayer. However the PCE$^{\text{SLME}}$ values are much smaller
than the ones obtained from SQ-limit, showing values lowers than
\SI{2}{\percent} independent of IPA or BSE levels, for all
monolayers. These huge difference can be attributed to the small
thickness of these monolayers, which consequently results in a lower
absorbance rate of these materials, which means that the material only
absorbs a small fraction of the incident photons. This characteristic
makes the material practically transparent, posing a significant
challenge for their applications in photovoltaic devices.

Indeed the application of light trapping techniques, as proposed by
Jariwala and co-workers,\cite{Jariwala_2962_2017} offers a promising
solution to enhance the absorbance of 2D materials, which can
potentially boost their solar harvesting performance, achieving an
absorbance rate closer to \SI{100}{\percent}. To address this
scenario, we considered the application of SLME method
(PCE$_{max}^{\text{SLME}}$) that approximates the absorbance curve, by
a Heaviside function, in the same way of SQ-limit, but also
considering the recombination fraction obtained in SLME from the
difference of direct and fundamental band gaps in IPA and excitonic
ground state and direct excitonic ground states, which lowers the PCE
when compared with SQ-limit results (PCE$^{\text{SQ}}$). In Group III
PCE$_{max}^{\text{SLME}}$ lies in the range of
\SIrange{15.70}{27.62}{\percent}(\SIrange{25.24}{32.23}{\percent}) at
BSE(IPA), Group IV
\SIrange{0.005}{8.02}{\percent}(\SIrange{7.87}{13.83}{\percent}) at
BSE(IPA) and Group V
\SIrange{11.71}{23.86}{\percent}(\SIrange{7.70}{14.99}{\percent}) at
BSE(IPA). The huge difference from PCE$_{max}^{\text{SLME}}$ and
PCE$^{\text{SQ}}$ are justified from the recombination fraction, as
for the majority of these systems the electronic band gap and
excitonic ground state are indirect, in group IV this was also
justified by the small or null absorption coefficient in parts of
visible and UV spectrum. \ce{InP3}, \ce{GaP3} and \ce{BiP3} shows a
good solar harvesting efficiency around
\SI{20}{\percent} - \SI{30}{\percent}, being attractive for solar cell applications.

\section{Conclusions}

In this work we use density-functional theory and molecular dynamics
simulations to investigate the thermal, electronic and optical
properties of triphosphide based two-dimensional materials
(XP$_3$). We find that with exception of InP$_3$, all structures have
indirect band gap. Furthermore, all systems show strong excitonic
effects. We show that mono-layered XP$_3$ exhibits optical absorption
with strong excitonic effects.  In particular, the exciton binding
energy is significantly large for X = Ga, Tl, Ge, Sn and Pb. Finally, AIMD calculations show that 2D-XP$_3$ is stable
 at room temperature, with exception of TlP$_3$ monolayer, which shows
 a strong distortion yielding to a phase separation of th\ e P and Tl
 layers. Finally, we show that  \ce{InP3}, \ce{GaP3} and \ce{BiP3} shows a good solar harvesting efficiency around \SI{20}{\percent} - \SI{30}{\percent}, being attractive for solar cell applications.

\section{Acknowledgments}
We acknowledge the financial support from the Brazilian funding agency CNPq under grant numbers 305174/2023-1, 313081/2017-4, 305335/2020-0, 309599/2021-0, 408144/2022-0, 305952/2023-4, 444069/2024-0 and 444431/2024-1. A.C.D also acknowledge FAPDF grants numbers 00193-00001817/2023-43 and 00193-00002073/2023-84. A.C.D and A.L.R also acknowledges DPG-FAPDF-CAPES Centro-Oeste grant 00193-00000867/2024-94. We thank computational resources from LaMCAD/UFG, Santos Dumont/LNCC, CENAPAD-SP/Unicamp (project number 897 and 761) and Lobo Carneiro HPC (project number 133).

\bibliography{refs}

%

\end{document}